\documentclass[prd,aps,11pt,nofootinbib, onecolumn,superscriptaddress,preprintnumbers,balancelastpage,longbibliography]{revtex4-1}

\usepackage{dcolumn} 
\usepackage{xcolor}
\usepackage{youngtab}
\usepackage{ytableau}

\usepackage{booktabs}    
\usepackage{placeins}
\usepackage{soul}

\definecolor{redd}{rgb}{0.8, 0.1,0.2}
\definecolor{navy}{rgb}{0.05, 0.23,0.75}
\usepackage[colorlinks=true,citecolor=red,linkcolor=blue]{hyperref}
\usepackage{chngcntr}
\hypersetup{
     colorlinks   = true,
     citecolor    = navy,
	linkcolor = redd,
	urlcolor=navy,
	anchorcolor=blue
}
\usepackage{bbm}
\usepackage{amsfonts}
\usepackage{amsmath,amssymb}
\usepackage{mathrsfs}
\usepackage{epsfig}
\usepackage{graphicx}
\usepackage{wrapfig}                 
\usepackage{url}
\usepackage{hyperref}
\usepackage{float}
\usepackage{color}
\usepackage{multirow}
\usepackage{lipsum}
\usepackage{enumitem}
\usepackage{tikz}
\usepackage{ctable}

\usepackage{enumitem}
\usepackage{chngcntr}
\usepackage{array} 

\usepackage[most]{tcolorbox}
\tcbuselibrary{theorems}
\tcbset{
  mytheobox/.style={
    colback=gray!10!white,
    colframe=black,
    boxrule=0.4pt,
    arc=1mm,
    fonttitle=\bfseries,
    title={#1},
    before skip=10pt,
    after skip=10pt,
    breakable,
    enhanced,
  }
}

\newcommand{\nn}{\nonumber}

\newcommand{\be}{\begin{equation}}
\newcommand{\ee}{\end{equation}}
\newcommand{\bea}{\begin{eqnarray}}
\newcommand{\eea}{\end{eqnarray}}
\newcommand{\bc}{\begin{center}}
\newcommand{\ec}{\end{center}}

\begin{document}
		
\title{
Spurion Analysis for Non-Invertible Selection Rules from Near-Group Fusions
}

\author{Motoo Suzuki}
\email{msuzuki@sissa.it}
\affiliation{SISSA International School for Advanced Studies, Via Bonomea 265, 34136, Trieste, Italy}
\affiliation{INFN, Sezione di Trieste, Via Valerio 2, 34127, Italy}
\affiliation{IFPU, Institute for Fundamental Physics of the Universe, Via Beirut 2, 34014 Trieste, Italy}

\author{Ling-Xiao Xu}
\email{lxu@ictp.it}
\affiliation{Abdus Salam International Centre for Theoretical Physics, Strada Costiera 11, 34151, Trieste, Italy}

\author{Hao Y. Zhang}
\email{haozhang@g.ecc.u-tokyo.ac.jp}
\affiliation{Kavli Institute for the Physics and Mathematics of the Universe (WPI), University of Tokyo, Kashiwa, Chiba 277-8583, Japan}

\begin{abstract}
We generalize the framework of spurion analysis to a class of selection rules arising from non-invertible fusion algebras in perturbation theory. As a first step toward systematic applications to particle physics, we analyze the near-group fusion algebras, defined by fusion rules built from a finite Abelian group $G$ extended by a single non-invertible element. Notable examples include the Fibonacci and Ising fusion rules.
We introduce a systematic scheme for labeling coupling constants at the level of the non-invertible fusion algebra, enabling consistent tracking of couplings when constructing composite amplitudes from simpler building blocks. Our labeling provides a clear interpretation of why the tree-level exact non-invertible selection rules are violated through radiative corrections, a unique phenomenon essential to ``loop-induced groupification''. We also identify the limit where the near-group fusion algebra is lifted to a $G\times \mathbb{Z}_2$ group, which provides an alternative scheme of spurion analysis consistent with the original one based on the near-group algebra. Meanwhile, we highlight the distinctions between the selection rules imposed by the near-group fusion algebra and those from breaking the $G\times \mathbb{Z}_2$ group.

\end{abstract}

\maketitle
\tableofcontents

\section{Introduction}
\label{sec:intro}

Symmetry serves as a powerful organizing principle in quantum field theories (QFTs). Over the past decade, the concept of global symmetries has been significantly generalized following the seminal work~\cite{Gaiotto:2014kfa}, which stimulated extensive cross-disciplinary research across high-energy physics, condensed matter physics, and mathematics, leading to a rich exchange of ideas and results. (For reviews from various perspectives, see e.g.~\cite{Cordova:2022ruw, McGreevy:2022oyu, Gomes:2023ahz, Schafer-Nameki:2023jdn, Brennan:2023mmt, Luo:2023ive, Shao:2023gho, Costa:2024wks, Iqbal:2024pee, Davighi:2025iyk}.) Among these generalizations, non-invertible symmetries~\cite{Bhardwaj:2017xup, Chang:2018iay} have emerged as a particularly striking development, highlighting that symmetry actions do not have to obey the group law.

Motivated by the developments of non-invertible symmetries, a class of selection rules from commutative non-invertible fusion algebras has been applied to study scattering amplitudes in perturbative $(3+1)$-dimensional QFTs, offering a potentially powerful framework for exploring physics beyond the Standard Model (SM). For clarity, we refer to them as \textit{non-invertible selection rules} (NISRs).
It was recently~\footnote{Non-group selection rules were studied long ago, however. See e.g.~\cite{Hamidi:1986vh, Font:1988nc, Kobayashi:1995py, Kobayashi:2011cw} for early seminal works.} shown in~\cite{Kaidi:2024wio} (see also \cite{Heckman:2024obe}) that NISRs are exact only at tree level, but are increasingly violated at higher loop orders, eventually reducing to a finite group. The whole procedure is called \emph{loop-induced groupification} of the NISRs.
Building on this observation, several NISRs have already been incorporated into particle physics models, primarily focusing on Yukawa texture zeros~\cite{Kobayashi:2024yqq, Kobayashi:2024cvp, Funakoshi:2024uvy, Kobayashi:2025znw, Liang:2025dkm, Kobayashi:2025ldi, Kobayashi:2025thd},  minimal extensions of the SM~\cite{Suzuki:2025oov}, models of radiative fermion masses~\cite{Kobayashi:2025cwx, Nomura:2025sod, Nomura:2025yoa, Chen:2025awz, Okada:2025kfm}, and discrete matter symmetries in supersymmetric models~\cite{Kobayashi:2025lar}. See also~\cite{Dong:2025jra} on constructing NISRs from discrete gauging (e.g., orbifolding).

From a particle physics perspective, it is natural to understand the violation of symmetries (or selection rules) using \emph{spurion analysis}, which involves promoting the coupling constants into (non-propagating) background fields that carry charges under the relevant symmetries. 
Those background fields are called \emph{spurions}.

Compared to ordinary (invertible) Abelian symmetry groups, the phenomenon of ``loop-induced groupification'' of NISRs, as we explained above, appears counterintuitive at first glance. As is well-known, selection rules from ordinary anomaly-free Abelian groups that hold at tree level remain exact to all loop orders in perturbation theory.~\footnote{Classical symmetries that are anomalous can be violated at the quantum level, where the breaking originates from the path integral measure rather than from any coupling in the Lagrangian.}
Indeed, when the Abelian group is exact at the tree level, all the tree couplings in the classical Lagrangian are labeled by the identity; hence, any other coupling labeled by a nontrivial group element cannot be generated through radiative corrections at any loop order in perturbation theory. 
Inspired by such reasoning, it is natural to label all the tree couplings by the identity element of the non-invertible fusion algebra when it is exact at the tree level. However, to formally preserve the NISRs at loop order, one must use the nontrivial elements to label the couplings that are radiatively generated while explicitly breaking the NISRs. From the naive perspective of spurion analysis, this seems contradictory: how can the couplings labeled by nontrivial elements arise from the fusion of those all labeled by the identity?

This apparent contradiction suggests that the framework of spurion analysis needs to be generalized when applied to NISRs.
Motivated by the examples studied in~\cite{Suzuki:2025oov}, we systematically generalize the spurion analysis to a class of NISRs arising from near-group fusion algebras~\cite{Evans:2012ta} in this paper.~\footnote{Ref.~\cite{Evans:2012ta} considers near-group fusion \emph{categories}, which supplement the fusion algebras by the additional data of the associators (i.e., F-symbols). Such data may be physically useful when resolving a 4-point vertex into a pair of 3-point vertices in different ways. We leave a study for future work.}
Since the ordering of particles in a vertex does not matter when determining whether the vertex is allowed, only commutative fusion algebras are physically relevant for our purposes. A near-group fusion algebra (which is commutative) contains the elements $\{g_i\}$ of a finite Abelian group $G$ extended by another non-invertible element $\rho$, whose fusion rules are given by
\bea
g_i \rho &=& \rho g_i =\rho\;, \label{eq:NG_1}\\
\rho^2 &=& n^\prime \rho+ \sum_{\text{all}\ g_i\in G} g_i \label{eq:NG_2} \;.
\eea
In addition, the elements $\{g_i\}$ obey the ordinary group law of $G$. From Eq.~\eqref{eq:NG_2}, the element $\rho$ is said to be self-conjugate, i.e., the fusion product of $\rho$ with itself contains the identity element. However, $\rho$ does not have an inverse, i.e. $\rho^2\neq \mathbbm{1}$ where $\mathbbm{1}$ is the identity. This is the hallmark of non-invertibility. 
Following~\cite{Evans:2012ta}, we denote the above non-invertible algebra as the type of $G+n^\prime$ with $n^\prime\in \mathbb{Z}_{\geq 0}$. For example, the Fibonacci fusion rules can be denoted as the $\{\mathbbm{1}\}+1$ type, and the Ising fusion rules can be denoted as the $\mathbb{Z}_2+0$ type; for example, see~\cite{Suzuki:2025oov} for some minimal particle physics models faithfully realizing these NISRs.

We now outline a procedure for performing the spurion analysis in the case of the non-invertible $G+n^\prime$ fusion algebra. We first need to promote coupling constants into background fields, and then label them using the elements in the fusion algebra.
The consistent labeling should allow us to achieve the following goals: 
\begin{itemize}
\item To systematically keep track of the couplings in all the scattering processes, both at tree and loop levels in perturbation theory. 
\item To reconcile the ``loop-induced groupification''~\cite{Kaidi:2024wio} of NISRs with the conventional wisdom of spurion analysis. 
\end{itemize}
These two goals are related, and they are realized by assigning nontrivial elements of the fusion algebra~\cite{Suzuki:2025oov} to coupling constants at tree level, even when the NISRs are not explicitly violated. This feature is distinct from the conventional spurion analysis for ordinary (invertible) Abelian symmetries, where all the couplings are labeled by the identity element when the symmetry group is exact. As we will easily see using spurion analysis, for the non-invertible fusion algebra of the $G+0$ type, there is a $\mathbb{Z}_2$ symmetry that remains exact up to all loop orders in perturbation theory. Such a result is consistent with~\cite{Kaidi:2024wio}. 

Furthermore, the consistent labeling of couplings implies that the non-invertible fusion rules in Eqs.~\eqref{eq:NG_1} and~\eqref{eq:NG_2} can be understood as a specific way of breaking a $G\times\mathbb{Z}_2$ group, which we call the \emph{lifted group}. Such a group is obtained once we switch off all the couplings with nontrivial labeling using the non-identity elements of the non-invertible $G+n^\prime$ fusion algebra, and it justifies that, when those couplings are suppressed, the suppression is technically natural in the sense of 't Hooft~\cite{tHooft:1979rat}. 
Accordingly, one can also relabel the coupling constants and dynamical fields using the elements of the lifted $G\times\mathbb{Z}_2$ group.
As we will show, the two labeling schemes --- the non-invertible $G+n^\prime$ fusion algebra versus the lifted $G\times\mathbb{Z}_2$ group --- are compatible, while being independent, thereby ensuring that our labeling procedure based on the non-invertible $G+n^\prime$ fusion algebra is self-consistent. 
Meanwhile, we note a crucial difference in the coupling patterns implied by the non-invertible $G+n^\prime$ fusion algebra compared to those arising from breaking the lifted $G\times\mathbb{Z}_2$ group. The former predicts a specific hierarchical structure among couplings distinguished by loop order, which cannot be accounted for within the latter framework.

At the conceptual level, NISRs should be viewed as a minimal and fundamental layer of structure even before organizing particles in complete representations of symmetry groups. After all, the representations of a group $\tilde{G}$ form a fusion algebra Rep$(\tilde{G})$ under the tensor product. While there is no first-principle reason that the most general particle interactions must be described by fusion algebras, we find them particularly natural because they are a fundamental layer of structure already existing in the ordinary representation theory of groups; it is therefore natural to consider their implications alone without imposing further structures. General non-invertible fusion algebras capture the same algebraic structure and can also be viewed as a generalization, since some basis elements can have non-integer quantum dimensions. However, since particles labeled by elements of a general fusion algebra are not in representations of symmetry groups, NISRs discard the structure of Clebsch-Gordan coefficients and invariant tensors that are used to contract indices of representations. In this sense, the usual requirement that particles are in representations of groups is algebraically more involved --- it requires tensor structures which are absent in NISRs. Precisely because NISRs involve less algebraic structure, they are less rigid under radiative corrections. The flip side of this observation is that \emph{completing particles into full representations of groups can be viewed as an obstruction to loop-induced groupification of NISRs}.
From a bottom-up perspective, NISRs in this paper are simply a labeling scheme for particles and couplings, and we do not assume that there is a symmetry transformation behind them. However, from a top-down perspective, we cannot exclude the possibility that NISRs emerge from symmetry actions in the ultraviolet. For instance, NISRs can be viewed as a low-energy field theory manifestation of non-invertible symmetries on the string worldsheet in perturbative string theory~\cite{Kaidi:2024wio, Heckman:2024obe, Hamidi:1986vh, Font:1988nc, Kobayashi:1995py, Kobayashi:2011cw}.

We view the results in this paper as a necessary step toward generalizations to other types of fusion algebras and more systematic applications to particle physics models. We notice that our reasoning for spurion analysis can potentially be generalized to other fusion algebras with more than one non-invertible element.
To illustrate this point, we will only discuss Conj$(S_3)$ as an example, while leaving a systematic generalization to future work~\cite{Suzuki:2025kxz, Xu:2026nwh}.

The rest of the paper is organized as follows. In Section~\ref{sec:sp}, we present our main general arguments of the generalized spurion analysis for NISRs from near-group fusions, with the aim of demonstrating that the goals outlined above are successfully achieved. In Section~\ref{sec:exp}, we explore a few simple examples illustrating the general framework developed in the previous section. In Section~\ref{sec:conc}, we briefly conclude and suggest some possible future directions. In Appendix~\ref{app_conj_s3}, we discuss a model based on the Conj$(S_3)$ fusion algebra, serving as an example beyond the class of near-group fusion algebras.

\section{Spurion Analysis: general theory}
\label{sec:sp}

In this section, we begin with a brief review of conventional spurion analysis, followed by our main result: its generalization to NISRs arising from near-group fusion algebras. Finally, we interpret the spurion analysis of the non-invertible $G+n^\prime$ fusion algebra from the perspective of the lifted $G\times\mathbb{Z}_2$ group, while highlighting the key distinctions between the two frameworks.

\subsection{Conventional spurion analysis for ordinary groups}
\label{sec:conv_sp}

In QFTs and particle physics, spurion analysis is a widely used method to systematically study symmetry-breaking effects. The central idea is to promote coupling constants to background non-dynamical fields, called spurions, which transform under a symmetry group so that the Lagrangian remains formally invariant.

For instance, consider a global symmetry described by an Abelian group $G$, and an interaction term 
\begin{equation}
    \lambda \ \phi_1 \dots \phi_k\;,
\end{equation}
where each field $\phi_i$ carries charge $g_i\in G$. If the product of these charges does not equal the identity element of $G$, i.e. $g_1 g_2\cdots g_k\neq 1$, the interaction term explicitly breaks the symmetry. In spurion analysis, one promotes the coupling constant $\lambda$ to a background non-dynamical field that carries the \emph{inverse} charge $c(\lambda)$ under the group $G$, 
\be
c(\lambda) = \left(g_1 g_2\cdots g_k\right)^{-1}\ .
\ee
This assignment ensures that the overall operator is neutral under $G$ and thus formally symmetric. The fact that the spurion carries the inverse of the total charges of the dynamical fields is a key feature of spurion analysis with \emph{invertible} Abelian symmetries, and it has to be contrasted with the non-invertible case discussed later. This procedure allows one to track symmetry-breaking effects in radiative corrections systematically and to identify the technically natural parameters in the sense of 't Hooft~\cite{tHooft:1979rat}: those whose vanishing restores a symmetry and are therefore protected from large quantum corrections. 

A classical example is provided by QED, where the electron mass term $m_e \bar{\psi}_L\psi_R+\text{h.c.}$ explicitly breaks the chiral $U(1)_L\times U(1)_R$ symmetry~\footnote{The axial part of the $U(1)_L\times U(1)_R$ symmetry suffers from the Adler-Bell-Jackiw anomaly, which, from a modern perspective, is understood as a non-invertible symmetry~\cite{Choi:2022jqy, Cordova:2022ieu}. However, for our interest in the current paper, it is still the Abelian $U(1)_L\times U(1)_R$ group that serves as the selection rules relevant for the helicity structure of the scattering processes in ordinary QED; see e.g. the scattering process $e^+e^-\to \mu^+\mu^-$ in~\cite{Peskin:1995ev}.}, under which the chiral fermions are charged as $\psi_L\sim (1,0)$ and $\psi_R\sim (0,1)$. By treating $m_e$ as a spurion with the charge $(1,-1)$, the mass term becomes formally invariant under the $U(1)_L\times U(1)_R$ group. In this way, the smallness of $m_e$ is technically natural, since taking $m_e\to 0$ restores the symmetry, which protects $m_e$ from quantum corrections. That is, if $m_e$ is the only parameter in the theory charged nontrivially under $U(1)_L\times U(1)_R$, it cannot be radiatively generated from other couplings that are all labeled by the identity element.

Similar reasoning applies in other areas of particle physics. For example, in chiral perturbation theory, quark masses are treated as spurions of $SU(3)_L\times SU(3)_R$~\cite{Weinberg:1968de}. In the framework of minimal flavor violation~\cite{DAmbrosio:2002vsn}, the Yukawa matrices play the same role under the SM flavor symmetry group $U(3)^5$; see e.g.~\cite{Grinstein:2024iyf} for a recent study.

\subsection{Generalized spurion analysis for NISRs from near-group fusion algebras}
\label{sec:gene_sp}

We now generalize the spurion analysis to NISRs arising from the $G+n^\prime$ fusion algebra.
We begin by constructing the most general classical Lagrangian consistent with the fusion algebra in Eqs.~\eqref{eq:NG_1} and~\eqref{eq:NG_2}. The construction follows these steps: 
\begin{itemize}
\item Each field $\phi_i$ in the theory is labeled by a basis element in the non-invertible $G+n^\prime$ fusion algebra.~\footnote{By ``basis elements'' we mean the irreducible elements of the fusion algebra --- analogous to irreducible representations of a group --- that cannot be decomposed further.} As indicated by the fusion rules in Eqs.~\eqref{eq:NG_1} and~\eqref{eq:NG_2}, each element has a conjugate, which can be used to label the conjugate field $\bar\phi_i$. This ensures that the identity appears in the fusion product of the elements labeling $\phi_i$ and $\bar\phi_i$, i.e., 
\be
\mathbbm{1}\prec c(\phi_i) \ c(\bar\phi_i)\;. 
\ee
Consequently, all $\phi_i$ are dynamical fields, which admit both kinetic and mass terms. 
\item Furthermore, an interaction term $\phi_i\phi_j\cdots \phi_l$ is allowed in the classical Lagrangian when the identity is contained in the fusion product of the elements labeling all the fields in the operator, i.e., 
\be
\mathbbm{1}\prec c(\phi_i) \ c(\phi_j)\ \cdots\  c(\phi_l)\;.  
\ee
All the interaction terms as above can be viewed as on-shell contact amplitudes~\footnote{The Lagrangian here serves only as a bookkeeping device that manifests the interactions allowed by the NISRs; the core of our discussion --- the labeling of couplings under the fusion algebra and their tracking through radiative corrections --- is formulated at the level of amplitudes and is unchanged even without reference to a Lagrangian.} involving the particles interpolated by the dynamical fields $\phi_i$.
\end{itemize}
Following Eqs.~\eqref{eq:NG_1} and~\eqref{eq:NG_2} and the above steps, we construct the classical Lagrangian for the $G+n^\prime$ fusion algebra with $n^\prime> 0$, including all the possible terms allowed by the algebra as follows:~\footnote{Notice that these terms coincide with the interaction terms between scalar particles without any other quantum numbers. Imposing other quantum numbers, such as spin or other internal symmetries, may set some of the couplings to zero.}
\begin{eqnarray}
\mathcal{L}_{n^\prime> 0}\supset && \lambda^{(0)}_{1} \mathbbm{1}\nonumber\\
&& + \lambda^{(0)}_{\rho^2} \rho^2 + \lambda^{(0)}_{g_ig_i^{-1}} g_i (g_i)^{-1} \nonumber\\
&& + \lambda^{(0)}_{g_ig_jg_k} g_i g_j g_k +  \lambda^{(0)}_{g_i\rho^2} g_i \rho^2 + \lambda^{(0)}_{\rho^3} \rho^3 \nonumber\\
&& + \lambda^{(0)}_{g_i g_j g_k g_l} g_i g_j g_k g_l + \lambda^{(0)}_{g_i g_j \rho^2} g_i g_j \rho^2 + \lambda^{(0)}_{g_i \rho^3} g_i \rho^3 + \lambda^{(0)}_{\rho^4} \rho^4 \;, 
\label{eq:cl_Lag}
\end{eqnarray}
where we have neglected all the kinetic terms and higher-dimensional operators, all the $\lambda$'s denote the coupling constants with the superscripts indicating their loop order. 
Here, we directly denote the particles by the basis elements that label them.~\footnote{Throughout the paper, we assume that the fusion algebra is realized faithfully by the particle content, i.e., each basis element labels at least one particle in the theory.}
For the interaction terms without the non-invertible element $\rho$, the elements $g_i$ have to satisfy the group law of $G$. In contrast, for the terms with $\rho$, the fusion algebra imposes no constraints on the allowed particles labeled by $g_i$. 
For the case of Tambara-Yamagami fusion rings, i.e., $n^\prime = 0$ as in Eq.~\eqref{eq:NG_2}, all the interaction terms with an odd power of $\rho$ fields in Eq.~\eqref{eq:cl_Lag} are forbidden, while the other terms are the same. (In particular, notice that all the terms linear in $\rho$ are always forbidden at tree level.)

Notice that Eq.~\eqref{eq:cl_Lag} includes only the tree-level couplings allowed by the NISRs imposed by Eqs.~\eqref{eq:NG_1} and~\eqref{eq:NG_2}. At the quantum level, additional couplings are generated through radiative corrections, which explicitly violate these selection rules.  
This occurs because when the particles labeled by the non-invertible element run in the loop, they open up new combinations of the external legs that were forbidden at the tree level, thereby violating the original NISRs. From a more formal viewpoint, such a procedure can be described by a quotient of the fusion ring viewed as a hypergroup~\footnote{A hypergroup is a generalization of a fusion algebra in which the product of two elements yields a formal sum of elements with coefficients that are not necessarily non-negative integers. See the appendix of~\cite{Kaidi:2024wio} for further details.}, where the quotiented out sub-hypergroup is generated by the elements appearing in the fusion product of conjugate pairs of non-invertible elements describing fields running in the loops. We refer the readers to~\cite{Kaidi:2024wio} for more details of this formal description. In the following, we describe a procedure of spurion analysis in the framework of NISRs, as partially discussed in~\cite{Suzuki:2025oov}.

Following the same logic of ordinary spurion analysis, we now promote the couplings in Eq.~\eqref{eq:cl_Lag} into background fields. 
We propose the following scheme to label the background fields at the level of non-invertible $G+n^\prime$ fusion algebra (with either $n^\prime>0$ or $n^\prime=0$): 
\begin{enumerate}
\item \label{rule1} If we replace all conjugate pairs of labels for external dynamical particles with the identity, then we demand that the remaining labels, from both the external background and dynamical fields, have their product that contains the identity.
\end{enumerate}
The replacement of conjugate pairs by the identity is physically motivated: a vertex without an additional conjugate pair can always be generated from the original vertex by closing a loop, so the two must carry the same spurion labeling.
This rule enables us to uniquely determine the labeling of background fields for NISRs derived from near-group fusion algebras. 
For example, the couplings in Eq.~\eqref{eq:cl_Lag} are labeled by the basis elements as
\begin{equation}
c(\lambda^{(0)}_{g_i\rho^2}) = (g_i)^{-1}, \quad c(\lambda^{(0)}_{\rho^3}) = \rho , \quad
c(\lambda^{(0)}_{g_i g_j \rho^2})  = (g_i g_j)^{-1}, \quad
c(\lambda^{(0)}_{g_i \rho^3}) = \rho\ .
\label{eq:portal_coup}
\end{equation}
All the other couplings are labeled with the identity. This is different from the conventional spurion analysis for Abelian symmetries reviewed in Section~\ref{sec:conv_sp}.
Here, certain couplings have nontrivial labelings even though the NISRs are exact at tree level.
We also see the difference in concrete examples in Section~\ref{sec:exp}.
This enables us to perform the spurion analysis \emph{independently} of the enhanced symmetry that is defined in the limit when any of the couplings in Eq.~\eqref{eq:cl_Lag} are switched off; our arguments work directly at the level of the $G+n^\prime$ fusion algebra.

One of the main advantages of the proposed labeling scheme is that it enables a systematic tracking of coupling constants when constructing composite amplitudes from simpler building blocks.~\footnote{Different from our scheme, one might otherwise label all the couplings in Eq.~\eqref{eq:cl_Lag} using the identity element since NISRs from the $G+n^\prime$ fusion algebra are exact in Eq.~\eqref{eq:cl_Lag}. However, as explained in Section~\ref{sec:intro}, such a labeling appears contradictory to the fact that NISRs-violating interactions can be radiatively generated.} We demonstrate this claim inductively, first at the tree level and then at the loop level.
Unlike the dynamical fields, a crucial feature of the spurions is that they are background fields that do not propagate. As a result, when constructing composite amplitudes from the simpler ones, only the dynamical fields can be glued together, while the background fields always remain as the external legs.~\footnote{In conventional spurion analysis, spurions are often introduced as vertex factors rather than external legs. Our treatment is equivalent: since spurions do not propagate, treating them as external legs simply assigns the identity as the vertex factor, while their quantum numbers are carried by the external leg. The key point is that spurion legs cannot be glued to form internal lines, which is the same constraint in either convention.} This provides a systematic treatment where each operator in the classical Lagrangian is viewed as a contact amplitude with external legs corresponding to both the dynamical and background fields. 

At the tree level, amplitudes can be constructed by gluing some dynamical fields, and the resulting amplitude is no longer a contact interaction. On the other hand, for any amplitude, there always exists a corresponding local operator that induces the same amplitude with the same set of external dynamical particles. 
\begin{enumerate}
\item Let us start with a general tree-level (not necessarily contact) amplitude $\mathcal{M}^{(0)}$, where the external dynamical particles are labeled by the elements in the non-invertible $G+n^\prime$ fusion algebra, i.e., 
\begin{equation}
\mathcal{M}^{(0)} = \lambda_{\mathcal{M}^{(0)}} \left(\rho \cdots \right) \left(g_i g_j \cdots \right).
\label{eq:amp0}
\end{equation}
where $\lambda_{\mathcal{M}^{(0)}}$ is the tree coupling, $(\rho ...)$ and $(g_i g_j ...)$ denote the non-invertible and invertible labels of the external particles, respectively.
Notice that, for a contact tree amplitude, $\lambda_{\mathcal{M}^{(0)}}$ is one of the couplings in the classical Lagrangian in Eq.~\eqref{eq:cl_Lag}. 
The labeling of $\lambda_{\mathcal{M}^{(0)}}$ is determined by the rule~\ref{rule1}.
When there are an even number of dynamical particles labeled by $\rho$, we can see that their fusion product contains the identity by grouping these (self-conjugate) elements into pairs. In this case, we have the labeling
\begin{equation}
c(\lambda_{\mathcal{M}^{(0)}})= \left(g_i g_j \cdots \right)^{-1}\;.
\end{equation} 
Otherwise, when the total number of external legs labeled by $\rho$ is odd, there would be an unpaired $\rho$. In this case, we label 
\begin{equation}
c(\lambda_{\mathcal{M}^{(0)}}) = \rho \;. 
\end{equation}
For convenience, we introduce the $\mathbb{Z}_2$-valued function $X(\rho; \mathcal{M})$ that denotes the number of unpaired $\rho$ fields in the amplitude $\mathcal{M}$. Clearly, $X(\rho; \mathcal{M}^{(0)})=0$ and $1$ for the two situations discussed above.

\item Now let us consider gluing two such tree-level amplitudes, $\mathcal{M}_1^{(0)}$ and $\mathcal{M}_2^{(0)}$, through a conjugate pair of dynamical particles with charges $c$ and $\bar{c}$. The resulting amplitude has external legs given by those of $\mathcal{M}_1^{(0)}$ and $\mathcal{M}_2^{(0)}$ excluding the conjugating pair: 
\begin{equation}
\mathcal{M}_1^{(0)}\underset{c}{\cup} \mathcal{M}_2^{(0)} = \lambda_{\mathcal{M}_1^{(0)}} \lambda_{\mathcal{M}_2^{(0)}} \left(\rho \cdots \right) \left(g_i g_j \cdots \right)\;,
\end{equation}
where $\left(g_i g_j \cdots \right)$ is the product of the remaining external legs in the two sub-amplitudes following the group law of $G$, while the number of unpaired $\rho$ in $\left(\rho \cdots \right)$ is given by
\begin{equation}
X(\rho; \mathcal{M}_1^{(0)}\underset{c}{\cup} \mathcal{M}_2^{(0)})= X(\rho; \mathcal{M}_1^{(0)})+X(\rho; \mathcal{M}_2^{(0)}) \quad \text{mod}\quad 2\;.
\end{equation}
From now on, we will suppress the label $c$, since our arguments would not depend on the precise charge of the state being glued---the function $X$ only counts the \emph{number} of unpaired $\rho$ labels, which is purely combinatorial.
As every amplitude corresponds to a local operator with the same external dynamical particles, we label the coupling for $\mathcal{M}_1^{(0)}\cup \mathcal{M}_2^{(0)}$ using the same rule~\ref{rule1}, i.e.,
\bea
c(\lambda_{\mathcal{M}_1^{(0)}\cup \mathcal{M}_2^{(0)}}) =  \left(g_i g_j \cdots \right)^{-1}\quad &\text{when}& \quad X(\rho; \mathcal{M}_1^{(0)}\cup \mathcal{M}_2^{(0)})=0\; ; \\
c(\lambda_{\mathcal{M}_1^{(0)}\cup \mathcal{M}_2^{(0)}}) =  \rho \quad &\text{when}& \quad X(\rho; \mathcal{M}_1^{(0)}\cup \mathcal{M}_2^{(0)})=1\ .
\eea
In this way, the labeling of the couplings is uniquely determined. In either case~\footnote{When $c(\lambda_{\mathcal{M}_1^{(0)}\cup \mathcal{M}_2^{(0)}})=g_k$, it implies either $X(\rho; \mathcal{M}_1^{(0)})=X(\rho; \mathcal{M}_2^{(0)})=0$ or $X(\rho; \mathcal{M}_1^{(0)})=X(\rho; \mathcal{M}_2^{(0)})=1$, Eq.~\eqref{spurion_1} takes the form of either $g_k \prec g_i g_j$ or $g_k\prec \rho \rho$, respectively; when $c(\lambda_{\mathcal{M}_1^{(0)}\cup \mathcal{M}_2^{(0)}})=\rho$, it implies either $X(\rho; \mathcal{M}_1^{(0)})=1$ (hence $X(\rho; \mathcal{M}_2^{(0)})=0$) or $X(\rho; \mathcal{M}_1^{(0)})=0$ (hence $X(\rho; \mathcal{M}_2^{(0)})=1$), Eq.~\eqref{spurion_1} takes the form of either $\rho \prec g_i \rho$ or $\rho \prec \rho g_j$, respectively.
As such, we verified the consistency of Eq.~\eqref{spurion_1} with the labeling rule~\ref{rule1} by explicitly working through all relevant cases in a brute-force manner. However, at a deeper level, such a consistency remains to be really understood. We hope to revisit this question in the future.}, we observe that the resulting coupling label satisfies
\begin{equation}
c(\lambda_{\mathcal{M}_1^{(0)}\cup \mathcal{M}_2^{(0)}}) \prec c(\lambda_{\mathcal{M}_1^{(0)}}) \ c(\lambda_{\mathcal{M}_2^{(0)}})\;, 
\label{spurion_1}
\end{equation}
i.e., the element labeling $\lambda_{\mathcal{M}_1^{(0)}\cup \mathcal{M}_2^{(0)}}$ is contained in the fusion product of the individual elements labeling $\lambda_{\mathcal{M}_1^{(0)}}$ and $\lambda_{\mathcal{M}_2^{(0)}}$.
\end{enumerate}
From Eq.~\eqref{spurion_1}, we see that the coupling $\lambda_{\mathcal{M}_1^{(0)}\cup \mathcal{M}_2^{(0)}}$ with any nontrivial labeling cannot be generated once all the couplings in Eq.~\eqref{eq:portal_coup} are switched off. For example, $\lambda_{\mathcal{M}_1^{(0)}\cup \mathcal{M}_2^{(0)}}$ with $c(\lambda_{\mathcal{M}_1^{(0)}\cup \mathcal{M}_2^{(0)}}) =  \rho$ cannot be generated without $\lambda^{(0)}_{\rho^3}$ or $\lambda^{(0)}_{g_i \rho^3}$ in the classical Lagrangian.

This completes our discussion on the spurion analysis for all tree-level amplitudes. Again, we emphasize that all the analysis is carried out at the level of the non-invertible $G+n^\prime$ fusion algebra. 
 
Finally, we apply the same rule~\ref{rule1} to determine the labeling of the coupling for loop amplitudes. 
The validity can be justified as follows:
\begin{enumerate}
\item Consider a loop amplitude at loop order $N$, denoted by $\mathcal{M}^{(N)}(k_1, \dots, k_m)$. By making $N$ cuts on the internal lines, the loop amplitude is reduced to a tree-level one, denoted by $\mathcal{M}^{(0)}(k_1, \dots, k_m; h_1, \overline{h}_1, \dots, h_N, \overline{h}_N)$, where the $k$'s and $h$'s are arbitrary elements in the fusion algebra, and the $h$'s are produced by cutting open the loops, with $h_i$ and $\overline{h}_i$ being a conjugate pair. 
Hence, the resulting tree amplitude has external legs that include the original external particles, along with $N$ additional conjugate pairs of dynamical particles corresponding to the internal lines that were cut.~\footnote{Notice that the arguments below also apply when making $i< N$ cuts, reducing the original $N$-loop amplitude $\mathcal{M}^{(N)}(k_1, \dots, k_m)$ to a $(N-i)$-loop amplitude $\mathcal{M}^{(N-i)}(k_1, \dots, k_m;  h_1, \overline{h}_1, \dots, h_i, \overline{h}_i)$. The resulting amplitude has external legs consisting of the original external particles, along with $i$ additional conjugate pairs of dynamical particles corresponding to the internal lines that were cut.}
\item According to the rule~\ref{rule1}, the labeling of the coupling depends only on the external legs that do not form conjugate pairs. Therefore, the $N$ conjugate pairs induced by the cuts do not affect the labeling. The remaining external legs in $\mathcal{M}^{(0)}(k_1, \dots, k_m; h_1, \overline{h}_1, \dots, h_N, \overline{h}_N)$ are identical to those of the original loop amplitude $\mathcal{M}^{(N)}(k_1, \dots, k_m)$, and the labeling of the corresponding coupling is preserved, i.e.,
\be
c\left(\lambda_{\mathcal{M}^{(N)}(k_1, \dots, k_m)}\right) = c\left(\lambda_{\mathcal{M}^{(0)}(k_1, \dots, k_m;  h_1, \overline{h}_1, \dots, h_N, \overline{h}_N)}\right)\ .
\label{spurion_1_plus}
\ee
In other words, the loop amplitude inherits its coupling labeling from the tree amplitude obtained by cutting open all its loops.
\item The results in Eqs.~\eqref{spurion_1} and~\eqref{spurion_1_plus} ensure that the basis element labeling the overall coupling of the loop amplitude that we analyze must be contained in the fusion product of the elements labeling each coupling involved in constructing the loop amplitude.
\end{enumerate}
As a result, an equation in the same form as Eq.~\eqref{spurion_1} holds true at the loop level in perturbation theory, where the precise value of the loop order depends on the number of conjugate pairs of dynamical particles that are glued together in the loops. 
This completes our discussion on the spurion analysis for all loop-level amplitudes.

The implications of the spurion analysis discussed above are as follows.
\begin{itemize}
\item For the couplings of the operators that are induced only at the loop level but forbidden at the tree level, they must be labeled by nontrivial elements of the fusion algebra. Consequently, if all the couplings in Eq.~\eqref{eq:portal_coup} are switched off, the process of ``loop-induced groupification''~\cite{Kaidi:2024wio} does not occur. 

\item It also implies that the tree-level couplings in Eq.~\eqref{eq:portal_coup} are technically natural in the sense of 't~Hooft~\cite{tHooft:1979rat}: switching them off enhances the symmetry to the lifted $G\times \mathbb{Z}_2$ group.~\footnote{As we will show in Section~\ref{sec_lifted_G}, the Lagrangian is invariant under a $G\times \mathbb{Z}_2$ symmetry in the limit where the couplings in Eq.~\eqref{eq:portal_coup} are switched off. This holds for all NISRs from near-group fusion algebras. An open question is how this conclusion extends to more general non-invertible fusion algebras, where a group responsible for technically natural parameters is not guaranteed to exist. See~\cite{Xu:2026nwh} for further discussions on this question.}
Furthermore, the couplings that are radiatively generated at loop level are also controlled by the fusion algebra structure: they inherit their nontrivial labeling from the tree-level portal couplings in Eq.~\eqref{eq:portal_coup}, and their smallness is dictated by the loop order at which they first appear, with the commutator of the fusion algebra being the relevant algebraic data~\cite{Kaidi:2024wio}. In particular, a loop-induced coupling vanishes up to a definite loop order dictated by the fusion algebra. This is analogous to, though distinct from, 't~Hooft's notion of technical naturalness, where setting a coupling to zero enhances a symmetry; here, the NISRs force a coupling to vanish up to a definite loop order without requiring an enhanced symmetry at that order.
\item Furthermore, for the scenario where $n^\prime=0$, the ``loop-induced groupification'' argument~\cite{Kaidi:2024wio} predicts an all-order exact $\mathbb{Z}_2$ symmetry, under which $\rho\to -\rho$ while all other elements are invariant. This is easily explained by the spurion analysis. Any amplitude involving an odd number of $\rho$ particles must be associated with a coupling labeled by $\rho$. However, when the fusion algebra is of the $G+0$ type (i.e., Tambara-Yamagami fusion algebras), none of the couplings in the classical Lagrangian is labeled by $\rho$. Therefore, the fusion rules in Eqs.~\eqref{eq:NG_1} and~\eqref{eq:NG_2} imply that such couplings labeled by $\rho$ can never be generated, preserving the $\mathbb{Z}_2$ symmetry at all loop orders. We note that while this conclusion can also be reached by inspecting the Lagrangian directly in simple cases, the spurion labeling provides a systematic and algorithmic approach that does not require examining the full Lagrangian --- an advantage that becomes significant for more complex fusion algebras where the number of allowed interaction terms can be large and the exact symmetry surviving at all loop orders is not easily read off by inspection.
\end{itemize}

\subsection{Grouplifting for NISRs from near-group fusion algebras}
\label{sec_lifted_G}

As we demonstrated in the last section, although the couplings in Eq.~\eqref{eq:portal_coup} are labeled nontrivially under the $G+n^\prime$ fusion algebra, the NISRs are exact at tree-level. 
However, these couplings are the sources of radiative violations of the NISRs at the quantum level.

Nevertheless, we observe that, once all the couplings in Eq.~\eqref{eq:portal_coup} with nontrivial labeling are switched off, the classical Lagrangian in Eq.~\eqref{eq:cl_Lag} reduces to
\begin{eqnarray}
\hat{\mathcal{L}}_{n^\prime> 0}\supset && \lambda^{(0)}_{1} \mathbbm{1}\nonumber\\
&& + \lambda^{(0)}_{\rho^2} \rho^2 + \lambda^{(0)}_{\rho^4} \rho^4  \nonumber\\
&& + \lambda^{(0)}_{g_ig_i^{-1}} g_i (g_i)^{-1} + \lambda^{(0)}_{g_ig_jg_k} g_i g_j g_k + \lambda^{(0)}_{g_i g_j g_k g_l} g_i g_j g_k g_l   \nonumber \\
&& + \lambda^{(0)}_{g_i g_i^{-1} \rho^2} \ g_i (g_i)^{-1} \rho^2 \; , 
\label{eq:cl_Lag_2}
\end{eqnarray}
which exhibits a lifted invertible symmetry group $G\times \mathbb{Z}_2$.  
For this reason, we refer to the couplings that are switched off above as the ``portal'' couplings. In their absence, the particle content governed by the non-invertible $G+n^\prime$ fusion algebra reorganizes into two sectors, each obeying an ordinary group law.

We note that the lifted group $G\times \mathbb{Z}_2$ is identified only in the limit when the portal couplings are switched off. However, these couplings are generally allowed by the NISRs but not by the lifted group --- this is precisely where the two frameworks disagree, and the NISRs should not be viewed as a substructure of the lifted group. Rather, the NISRs impose a restricted breaking pattern with concrete physical consequences that do not arise in generic symmetry-breaking scenarios.
For instance, a broken $\mathbb{Z}_2$ group alone cannot explain the absence of all the interaction terms linear in the particle labeled by $\rho$ --- such as $g_i \rho$ and $g_i g_j \rho$ --- while allowing the terms involving any other odd powers of $\rho$ --- such as $\rho^3$ and $g_i\rho^3$ --- to appear at tree level, as seen in the classical Lagrangian of Eq.~\eqref{eq:cl_Lag}.
However, this pattern is naturally explained by the NISRs from the $G+n^\prime$ fusion algebra with $n^\prime>0$. As a result, the interactions such as $g_i \rho$ and $g_i g_j \rho$ are predicted to be suppressed by a one-loop factor compared to $\rho^3$ and $g_i\rho^3$.
Moreover, as seen in Eq.~\eqref{eq:cl_Lag}, the breaking of the group law of $G$ originates from the tree-level interactions involving the $\rho$ field, which in turn induce $G$-breaking interactions that do not involve $\rho$ at one-loop order. In contrast, from the conventional viewpoint of a broken $G$ group, all the symmetry-breaking terms are on the same footing and thus expected to appear at the same loop order.
In summary, the NISRs can be understood as a version of the lifted group that is broken in a highly structured way, and their predictive power lies in this restricted breaking pattern and the hierarchical emergence of different types of couplings through radiative corrections. The precise algebraic characterization of this structure remains an interesting open question; some related discussions are further explored in~\cite{Xu:2026nwh}.

Despite these differences, it is natural to also perform the spurion analysis at the level of the lifted $G\times \mathbb{Z}_2$ group, following the conventional arguments reviewed in Section~\ref{sec:conv_sp}. 
To do so, we need to relabel all the dynamical fields and coupling constants using the charges (i.e., representations) of $G\times \mathbb{Z}_2$.~\footnote{Strictly speaking, in standard spurion analysis, fields and couplings are labeled by charges (or more generally, representations) of the symmetry group, not by group elements. However, since $G$ is a finite Abelian group in all our examples, the group of characters (i.e., the Pontryagin dual $\hat{G}$) is isomorphic to $G$ itself. We therefore use group elements and charges interchangeably throughout the paper.}
We define a surjective map 
\be
\varphi: \quad G\times \mathbb{Z}_2 \longmapsto G+n^\prime\;,
\ee
such that the elements of $G+n^\prime$ are obtained as the images of the elements in $G\times \mathbb{Z}_2$. Explicitly, we define  
\be
\varphi \left[(g_i, \hat{\rho})\right] = \rho \;, \quad\quad
\varphi \left[(g_i, 1)\right] = g_i\;, 
\label{eq:map}
\ee
for any $g_i\in G$. Here, the elements $g_i$ obey the group law of $G$, and $\hat{\rho}$ satisfies the group law of a $\mathbb{Z}_2$ group, i.e., $\hat{\rho}^2=1$.
For the dynamical particles in Eq.~\eqref{eq:cl_Lag}, we relabel the particle $\rho$ using the group element $(1, \hat{\rho}) \in G\times \mathbb{Z}_2$, and the particles $g_i$ using $(g_i, 1) \in G\times \mathbb{Z}_2 $.~\footnote{Notice that such a relabeling is not unique in some cases. For instance, one might alternatively relabel the $\rho$ particle using the group element $(-1, \hat{\rho}) \in G\times \mathbb{Z}_2$ when $G$ contains a $\mathbb{Z}_2$ subgroup. Accordingly, this modifies the labeling of couplings in Eq.~\eqref{eq:lift_lab} when the power of $\rho$ involved in the couplings is odd. Nonetheless, Eq.~\eqref{spurion_2} exactly holds the same. }
Following the conventional spurion analysis, the couplings in the Lagrangian are labeled by the inverses of the total charges of the dynamical fields. For instance, the couplings in Eq.~\eqref{eq:portal_coup} are
\begin{align}
\begin{split}
c(\lambda^{(0)}_{g_i\rho^2}) &= \left((g_i)^{-1}, 1\right), \quad c(\lambda^{(0)}_{\rho^3})= \left(1, \hat{\rho} \right) , \quad \\
c(\lambda^{(0)}_{g_i g_j \rho^2}) &= \left((g_i g_j)^{-1}, 1\right), \quad
c(\lambda^{(0)}_{g_i \rho^3})= \left(g_i^{-1}, \hat{\rho} \right)\ ,
\label{eq:lift_lab}
\end{split}
\end{align}
while all the other couplings in Eq.~\eqref{eq:cl_Lag} are labeled by $(1,1)$ at the level of the lifted $G\times \mathbb{Z}_2$ group.
Notice that the dynamical fields are not charged under either $G$ or $\mathbb{Z}_2$, while the background fields carry charges under both. Furthermore, we note that the labeling in Eq.~\eqref{eq:lift_lab}, together with the map defined in Eq.~\eqref{eq:map}, is consistent with the original non-invertible labeling of the portal couplings in Eq.~\eqref{eq:portal_coup}.

More generally, the coupling for any amplitude is labeled by the inverse of the product of the group elements associated with the other external dynamical particles. Using the invertibility of the group law, one can show inductively --- following the same reasoning as in the non-invertible case --- that the composition of couplings satisfies
\begin{equation}
c(\lambda_{\mathcal{M}_1 \cup \mathcal{M}_2}) = c(\lambda_{\mathcal{M}_1}) \  c(\lambda_{\mathcal{M}_2}) \ ,
\label{spurion_2}
\end{equation}
which is the analog of Eq.~\eqref{spurion_1}, but also valid for loop amplitudes. The map defined in Eq.~\eqref{eq:map} ensures the mutual consistency between the two labeling schemes, one based on the non-invertible $G+n^\prime$ fusion algebra and the other on the lifted $G\times \mathbb{Z}_2$ group.

While the results are consistent, one still needs to distinguish the spurion analysis at the level of the lifted $G \times \mathbb{Z}_2$ from that at the level of $G+n^\prime$ non-invertible fusion algebra. 
Conceptually, the former resembles the case for ordinary symmetries, where spurion analysis must be performed with respect to a broken group, since otherwise all couplings are charged as singlets if the symmetry group is exact. However, in the latter, couplings carry nontrivial labels already at the level of the classical Lagrangian, where the NISRs are exact. Hence, spurion analysis can be performed directly without relying on any symmetry enhancement.

\section{Examples of NISRs from near-group fusion algebras}
\label{sec:exp}

In this section, we present several concrete examples to illustrate the general framework developed in Section~\ref{sec:sp}.
For simplicity, we consider models only consisting of scalar particles, and we further assume that these scalars carry no additional quantum numbers. After all, introducing spin or any other quantum numbers amounts to imposing additional selection rules, which may set certain coupling constants to zero. However, the essential structure of the analysis remains the same.

The fusion algebras studied below are partially motivated by the particle physics models in~\cite{Suzuki:2025oov}, which are arguably the simplest models beyond the SM faithfully realizing the NISRs. Other near-group fusion algebras can be found in~\cite{Evans:2012ta}.

\subsection{Fibonacci fusion algebra}
\label{subsec:fib}
We begin with the Fibonacci fusion algebra (Fib) as the simplest example. In the notation of $G+n^\prime$ fusion algebra, it corresponds to the type $\{\mathbbm{1}\}+1$. In addition to the trivial group consisting of only the identity element $\mathbbm{1}$, we have the fusion rules for the non-invertible element $\tau$ as
\be
\mathbbm{1} \tau=\tau \mathbbm{1}=\tau, \quad\quad \tau^2= \mathbbm{1} + \tau\;,
\label{eq:fib}
\ee
which is an example of Eqs.~\eqref{eq:NG_1} and~\eqref{eq:NG_2}.

The simplest model that faithfully realizes the Fib consists of two real scalar particles, which are labeled as $\phi_1 \sim \mathbbm{1}$ and $\phi_2\sim \tau $. We consider the following classical Lagrangian exact under the Fib:
\begin{eqnarray}
\mathcal{L}_{\text{Fib}}\supset && \lambda^{(0)}_{1} \mathbbm{1}\nonumber\\
&& + \lambda^{(0)}_{\tau^2} \tau^2 + \lambda^{(0)}_{1^2} \mathbbm{1}^2 \nonumber\\
&& + \lambda^{(0)}_{1^3} \mathbbm{1}^3 +  \lambda^{(0)}_{1\tau^2} \mathbbm{1} \tau^2 + \lambda^{(0)}_{\tau^3} \tau^3 \nonumber\\
&& + \lambda^{(0)}_{1^4} \mathbbm{1}^4 + \lambda^{(0)}_{1^2 \tau^2} \mathbbm{1}^2 \tau^2 + \lambda^{(0)}_{1\tau^3} \mathbbm{1} \tau^3 + \lambda^{(0)}_{\tau^4} \tau^4 \; ,
\label{eq:cl_Lag_Fib}
\end{eqnarray}
which is an example of Eq.~\eqref{eq:cl_Lag}.
The crucial feature of $\mathcal{L}_{\text{Fib}}$ is that all the terms linear in $\tau$ are forbidden by the fusion rules at tree level, while all higher powers of $\tau$ are allowed. As a result, all the single-$\phi_2$ scattering processes must be suppressed by a loop factor relative to the other processes involving multiple $\phi_2$ particles.  
This hierarchical structure provides a potential experimental signature to test the Fib; see~\cite{Suzuki:2025oov} for related discussions in the context of a model extending the SM with a real scalar.  

Following our labeling rule~\ref{rule1}, the portal couplings in Eq.~\eqref{eq:cl_Lag_Fib} are labeled by the non-invertible element $\tau$, i.e.,
\begin{equation}
c(\lambda^{(0)}_{\tau^3})= \tau , \quad
c(\lambda^{(0)}_{1\tau^3} )= \tau\ .
\label{eq:fib_coup}
\end{equation}
All the other couplings are labeled with the identity. Although $\mathcal{L}_{\text{Fib}}$ fully respects the Fib, the couplings $\lambda^{(0)}_{\tau^3}$ and $\lambda^{(0)}_{1\tau^3}$ are still labeled nontrivially. 
It implies that radiative corrections involving these couplings can generate amplitudes linear in $\tau$, thereby violating the Fib at loop level. Such a result is consistent with the arguments in~\cite{Kaidi:2024wio}.

\begin{figure}[t]
\centering
\includegraphics[scale=0.22]{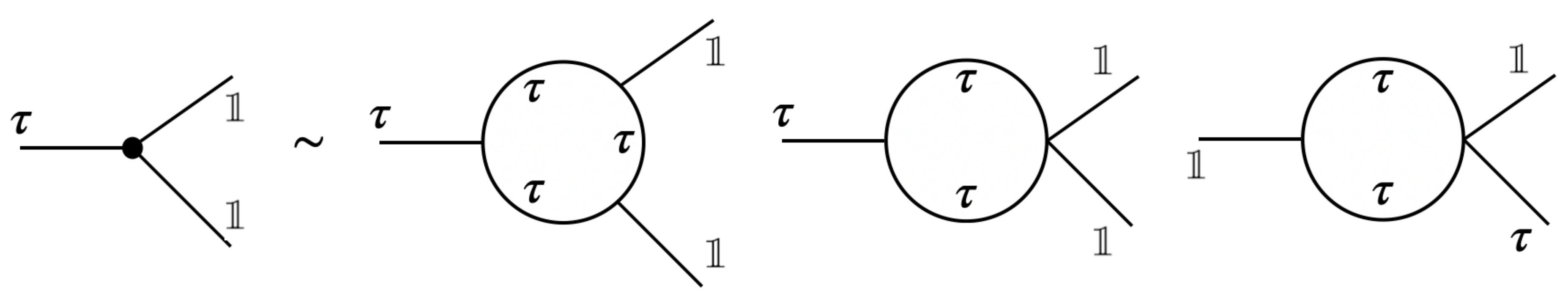}
\caption{A cartoon illustrating the one-loop scattering process $\phi_2\to \phi_1^2$ with the Fibonacci fusion algebra, where the dynamical particle $\phi_1$ is labeled by $\mathbbm{1}$ while $\phi_2$ is labeled by $\tau$. }
\label{fig:fib1}
\end{figure}

Following the spurion analysis in Section~\ref{sec:sp}, the labeling in Eq.~\eqref{eq:fib_coup} enables us to systematically track the couplings $\lambda^{(0)}_{\tau^3}$ and $\lambda^{(0)}_{1\tau^3}$ in all the scattering processes. In particular, we can test the general arguments in Section~\ref{sec:sp} by constructing complex tree and loop amplitudes involving these couplings. 
For example, we consider the scattering process $\phi_2\to \phi_1^2$, corresponding to the operator labeled by the basis elements $\tau \mathbbm{1}^2$. Since such an operator is inconsistent with the Fibonacci fusion rules in Eq.~\eqref{eq:fib}, so the process vanishes exactly at tree level but is generated at one-loop level. According to the labeling rule~\ref{rule1}, the associated coupling is labeled as
\begin{equation}
c(\lambda^{(1)}_{\tau 1^2})= \tau\; ,
\end{equation}
where the superscript indicates the loop order. 
As shown in Fig.~\ref{fig:fib1}, the one-loop amplitude can be built from tree-level interactions in Eq.~\eqref{eq:cl_Lag_Fib}, leading to the following relation between the couplings 
\be
\lambda^{(1)}_{\tau 1^2} = \frac{1}{16 \pi^2} \left( \lambda^{(0)}_{\tau^3} \left[\lambda^{(0)}_{1\tau^2}\right]^2+\lambda^{(0)}_{\tau^3} \lambda^{(0)}_{1^2\tau^2} + \lambda^{(0)}_{1\tau^2} \lambda^{(0)}_{1\tau^3}\right)\;,
\label{eq:spurion_fib}
\ee
where $\frac{1}{16 \pi^2}$ specifies the loop order where the coupling $\lambda^{(1)}_{\tau 1^2}$ is generated. We find that 
Eq.~\eqref{eq:spurion_fib} is consistent with the algebraic relation at the level of Fib
\begin{equation}
\tau\prec \tau \mathbbm{1}^2, \quad \tau\prec \tau \mathbbm{1} , \quad \tau\prec\mathbbm{1} \tau\ . 
\end{equation}
The key observation is that $\lambda^{(1)}_{\tau 1^2}$ is not generated in radiative corrections if both the portal couplings in Eq.~\eqref{eq:fib_coup} are switched off. 
Consequently, if the couplings in Eq.~\eqref{eq:fib_coup} are suppressed, their suppression is technically natural.

In the limit where $\lambda^{(0)}_{\tau^3}$ and $\lambda^{(0)}_{1\tau^3}$ are switched off, the theory is invariant under the lifted $\{\mathbbm{1}\} \times \mathbbm{Z}_2$ group. We define the surjective map (after suppressing the trivial group $\{\mathbbm{1}\}$)
\be
\varphi: \quad  \mathbb{Z}_2 \longmapsto \text{Fib}\;,
\ee
such that the elements of Fib are obtained as the images of the elements in $\mathbb{Z}_2$. Following Eq.~\eqref{eq:map}, we define
\be
\varphi \left[\hat{\tau}\right] = \tau \;, \quad\quad
\varphi \left[1\right] = \mathbbm{1} \;,
\ee
where $\hat{\tau}$ obey the group law of the $\mathbbm{Z}_2$ group, i.e., $\hat\tau^2=1$. It is natural to relabel all the dynamical fields and couplings using $\mathbbm{Z}_2$ group elements, e.g., 
\be
c(\lambda^{(0)}_{\tau^3}) = \hat\tau , \quad c(\lambda^{(0)}_{1\tau^3}) = \hat\tau , \quad c(\lambda^{(1)}_{\tau 1^2}) = \hat\tau\; .
\ee
The consistency of the fusion labels in Eq.~\eqref{eq:spurion_fib} is then manifest from the group law of the lifted $\mathbbm{Z}_2$ group, i.e., 
\begin{equation}
\hat\tau = \hat\tau 1^2, \quad  \hat\tau =\hat \tau 1, \quad \hat\tau = 1 \hat\tau \;.
\end{equation}
Thus, the two labeling schemes based on Fib and the lifted $\mathbbm{Z}_2$ group are mutually consistent. However, we emphasize that, while NISRs from Fib can be viewed as a special case of those from a broken $\mathbbm{Z}_2$ group, the hierarchical structure of the couplings discussed after Eq.~\eqref{eq:cl_Lag_Fib} cannot be explained using generic considerations of a broken $\mathbb{Z}_2$ group.

The Fibonacci fusion rule case is our simplest nontrivial example of NISRs from near-group fusion algebras. Beyond the scattering process discussed above, one can explore other scattering amplitudes involving additional external particles and at higher loop orders, and verify our arguments in Section~\ref{sec:gene_sp}.

\subsection{Tambara-Yamagami fusion algebra of $\mathbb{Z}_N$}
\label{subsec:TY}

\begin{figure}[t]
\centering
\includegraphics[scale=0.20]{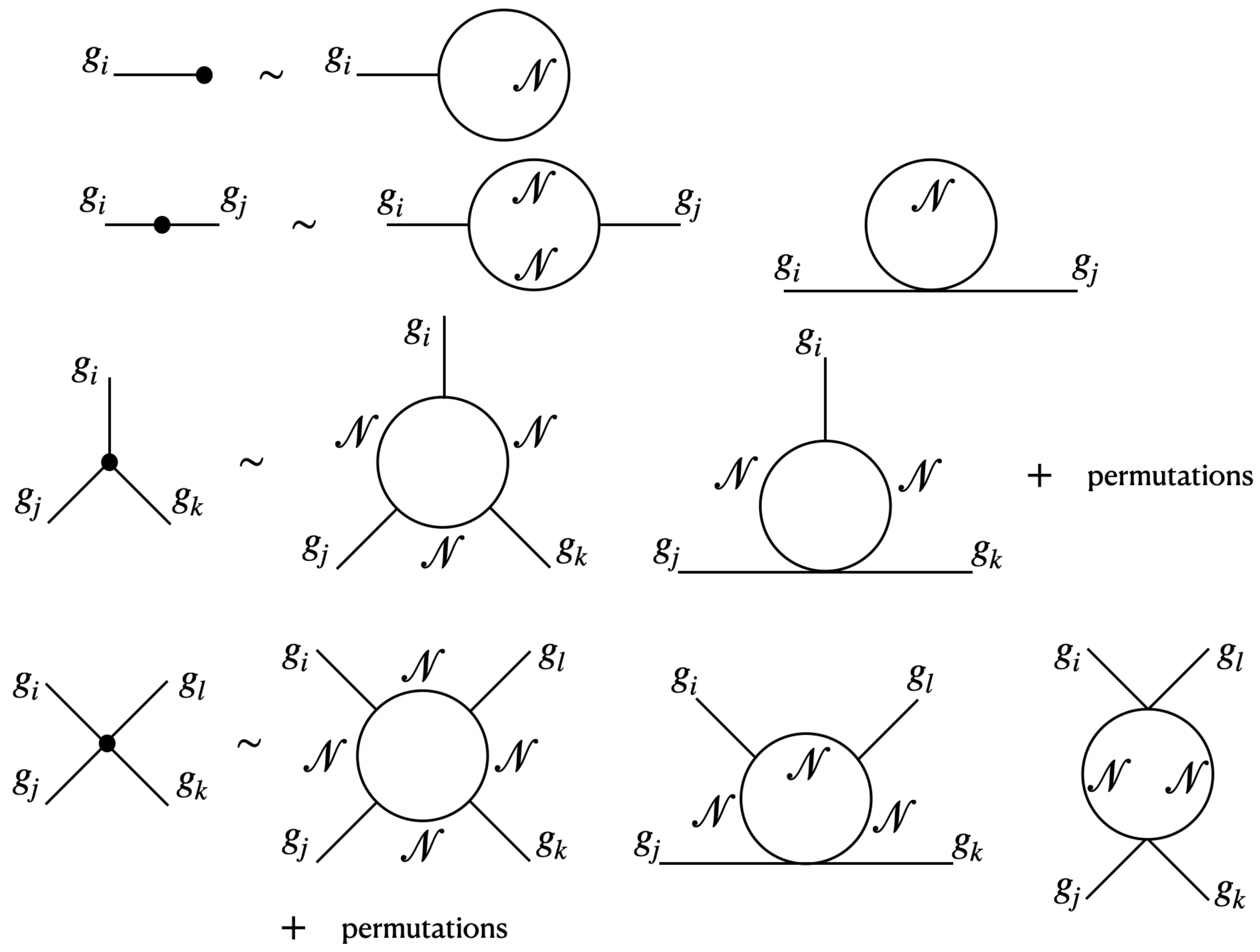}
\caption{A cartoon illustrating the one-loop scattering processes that correspond to the operators forbidden by TY$(\mathbb{Z}_N)$ at tree level. As induced by the loop of $\mathcal{N}$, the external legs $g_i$ do not have to obey the group law of a $\mathbb{Z}_N$ group at one-loop level.
}
\label{fig:TY}
\end{figure}

Next, we consider the Tambara-Yamagami (TY) fusion algebra associated with the Abelian group $\mathbb{Z}_N$, denoted as TY$(\mathbb{Z}_N)$. Notice that TY$(\mathbb{Z}_2)$ coincides with the Ising fusion algebra. 
In the notation of $G+n^\prime$ near-group fusion algebras, TY$(\mathbb{Z}_N)$ corresponds to the type $\mathbb{Z}_N+0$.
Apart from the elements $g_\ell=e^{\frac{2\pi i \ell}{N}}$ ($\ell=0,1,\cdots,N-1$) satisfying the group law of a $\mathbb{Z}_N$ group,%
\footnote{For $\mathbb{Z}_N~(N>2)$, the dynamical particles labeled by the elements $g_i$ are in general complex.}
TY$(\mathbb{Z}_N)$ includes the fusion rules for the non-invertible element $\mathcal{N}$ as
\bea
g_\ell \mathcal{N} = \mathcal{N} g_\ell =\mathcal{N}\;, \quad \mathcal{N}^2 = \sum_{ \text{all} \ g_\ell \in \mathbb{Z}_N } g_\ell \label{eq:TY} \; ,
\label{eq:TY_ZN_fusion_rules}
\eea
which are consistent with Eqs.~\eqref{eq:NG_1} and~\eqref{eq:NG_2}. The classical Lagrangian that faithfully realizes the TY$(\mathbb{Z}_N)$ is
\begin{eqnarray}
\mathcal{L}_{\text{TY}(\mathbb{Z}_N)}\supset && \lambda^{(0)}_{1} \mathbbm{1} \label{eq:Lag_TY} \\
&& + \lambda^{(0)}_{\mathcal{N}^2} \mathcal{N}^2 + \sum_i \lambda^{(0)}_{g_ig_i^{-1}} g_i (g_i)^{-1} \nonumber\\
&& + \sum_{i,j,k} \lambda^{(0)}_{g_i g_j g_k} \ g_i g_j g_k \ \delta_{(i + j + k) \bmod N,\, 0} +  \sum_i \lambda^{(0)}_{g_i\mathcal{N}^2} \ g_i \mathcal{N}^2  \nonumber\\
&& + \sum_{i,j,k,l} \lambda^{(0)}_{g_i g_j g_k g_l} \ g_i g_j g_k g_l \ \delta_{(i + j + k + l) \bmod N,\, 0} + \sum_{i,j} \lambda^{(0)}_{g_i g_j \mathcal{N}^2} \ g_i g_j \mathcal{N}^2 + \lambda^{(0)}_{\mathcal{N}^4} \ \mathcal{N}^4 \;, \nonumber
\end{eqnarray}
where we use the basis elements to denote the corresponding particles that they label.
Here are some interesting features of Eq.~\eqref{eq:Lag_TY}. 
\begin{itemize}
\item Only the terms with even powers of the non-invertible element $\mathcal{N}$ are allowed. Hence, there is a global $\mathbbm{Z}_2$ symmetry group arising from the ``loop-induced groupification'' of the TY$(\mathbb{Z}_N)$~\cite{Kaidi:2024wio}, under which $\mathcal{N}\to -\mathcal{N}$ while $g_i \in \mathbb{Z}_N$ are invariant. 
\item For the interaction terms without involving $\mathcal{N}$, the elements $g_i$ obey the group law of an ordinary $\mathbbm{Z}_N$ group at tree level. 
\item In contrast, for the terms involving $\mathcal{N}$, such as $g_i \mathcal{N}^2$ and $g_i g_j \mathcal{N}^2$, the $g_i$ are unconstrained and do not need to satisfy the group law of $\mathbbm{Z}_N$, since different $\mathbbm{Z}_N$ charges become indistinguishable after multiplying with $\mathcal{N}^2$.
In turn, the interaction terms involving $\mathcal{N}$ can induce new interactions among the $g_i$ at one-loop level via $\mathcal{N}$ loops; see in Fig.~\ref{fig:TY}. These radiatively generated terms generally violate the $\mathbb{Z}_N$ group law.
\end{itemize}
This hierarchical structure between the couplings of the $\mathbb{Z}_N$-violating interactions, such that interactions without $\mathcal{N}$ come at a higher loop order than interactions with $\mathcal{N}$, is a hallmark of the NISRs from TY$(\mathbb{Z}_N)$. In contrast, for the selection rules arising from a broken $\mathbb{Z}_N$ group, all symmetry-breaking terms are expected to appear at the same loop order.

Following our labeling rule~\ref{rule1}, the portal couplings are labeled as
\be
c(\lambda^{(0)}_{g_i\mathcal{N}^2})= g_i^{-1}\;,\quad  c(\lambda^{(0)}_{g_i g_j\mathcal{N}^2})= g_i^{-1} g_j^{-1}\ .
\label{eq:TY_coup}
\ee
All the other couplings are labeled by the identity.%
\footnote{Notice that $\lambda^{(0)}_{g_i\mathcal{N}^2}$ and $\lambda^{(0)}_{g_i g_j\mathcal{N}^2}$ are in general complex couplings for $\mathbb{Z}_N$ with $N>2$.}
Although $\mathcal{L}_{\text{TY}(\mathbb{Z}_N)}$ is exact under the $\text{TY}(\mathbb{Z}_N)$, the couplings in Eq.~\eqref{eq:TY_coup} are labeled nontrivially, which is again a hallmark of spurion analysis under NISRs. 
This labeling allows us to immediately understand the all-order exact $\mathbb{Z}_2$ group under which $\mathcal{N}\to -\mathcal{N}$, as suggested by~\cite{Kaidi:2024wio}: if any amplitude with an odd number of $\mathcal{N}$ particles were generated radiatively, its overall coupling would be labeled by $\mathcal{N}$. However, the element $\mathcal{N}$ cannot be obtained from the fusion product of any powers of the couplings in Eq.~\eqref{eq:TY_coup}. In other words, our labeling scheme above explains why the operators with odd number of $\mathcal{N}$, such as $\mathcal{N}$, $\mathcal{N} g_i$, $\mathcal{N} g_i g_j$, $\mathcal{N}^3$, $\mathcal{N} g_i g_j g_k$, $\mathcal{N}^3 g_i$, are not generated in radiative corrections at any loop order.

The labeling in Eq.~\eqref{eq:TY_coup} enables us to systematically track these couplings in scattering processes.
Let us consider one-loop amplitudes with the external particles corresponding to the operators like $g_i$, $g_i g_j$ with $i+j\neq 0$ mod $N$, $g_i g_j g_k$ with $i+j+k\neq 0$ mod $N$, $g_i g_j g_k g_l$ with $i+j+k+l\neq 0$ mod $N$. These operators are forbidden in the classical Lagrangian in Eq.~\eqref{eq:Lag_TY}, but the corresponding amplitudes are generated radiatively. According to our labeling rule~\ref{rule1}, their couplings are labeled as 
\be
c(\lambda^{(1)}_{g_i})= g_i^{-1}, \quad 
c(\lambda^{(1)}_{g_i g_j})= g_i^{-1} g_j^{-1}, \quad c(\lambda^{(1)}_{g_i g_j g_k})= g_i^{-1} g_j^{-1} g_k^{-1}, \quad 
c(\lambda^{(1)}_{g_i g_j g_k g_l})= g_i^{-1} g_j^{-1} g_k^{-1} g_l^{-1} \;.
\label{eq:TY_coup_2}
\ee
As illustrated in Fig.~\ref{fig:TY}, these one-loop amplitudes can be constructed from the tree vertices in Eq.~\eqref{eq:Lag_TY}, leading to the relation between couplings such as
\bea
\lambda^{(1)}_{g_i} & = & \frac{1}{16 \pi^2} \lambda^{(0)}_{g_i \mathcal{N}^2},\\
\lambda^{(1)}_{g_i g_j} & = & \frac{1}{16 \pi^2} \left(\lambda^{(0)}_{g_i\mathcal{N}^2} \lambda^{(0)}_{g_j\mathcal{N}^2} + \lambda^{(0)}_{g_i g_j\mathcal{N}^2} \right),\\
\lambda^{(1)}_{g_i g_j g_k} & = & \frac{1}{16 \pi^2} \left( \lambda^{(0)}_{g_i\mathcal{N}^2} \lambda^{(0)}_{g_j\mathcal{N}^2} \lambda^{(0)}_{g_k\mathcal{N}^2} + \lambda^{(0)}_{g_i\mathcal{N}^2} \lambda^{(0)}_{g_j g_k\mathcal{N}^2} + \text{permutations} \right),\\
\lambda^{(1)}_{g_i g_j g_k g_l} & = & \frac{1}{16 \pi^2} \left( \lambda^{(0)}_{g_i\mathcal{N}^2} \lambda^{(0)}_{g_j\mathcal{N}^2} \lambda^{(0)}_{g_k\mathcal{N}^2} \lambda^{(0)}_{g_l\mathcal{N}^2} + \lambda^{(0)}_{g_i\mathcal{N}^2} \lambda^{(0)}_{g_l\mathcal{N}^2} \lambda^{(0)}_{g_j g_k\mathcal{N}^2} + \lambda^{(0)}_{g_i g_l\mathcal{N}^2} \lambda^{(0)}_{g_j g_k\mathcal{N}^2} + \text{permutations} \right). \nn \\
\eea
These relations are consistent with the labeling at the level TY$(\mathbb{Z}_N)$, i.e., 
\bea
g_i^{-1} & \prec & g_i^{-1},\\
(g_i g_j)^{-1} & \prec &  g_i^{-1} g_j^{-1} ,\quad (g_i g_j)^{-1} \prec (g_i g_j)^{-1} ,\\
(g_i g_j g_k)^{-1} & \prec &   g_i^{-1} g_j^{-1} g_k^{-1} , \quad (g_i g_j g_k)^{-1} \prec g_i^{-1} (g_j g_k)^{-1}, \quad \cdots ,\\
(g_i g_j g_k g_l)^{-1} & \prec &  g_i^{-1} g_j^{-1} g_k^{-1} g_l^{-1}, \quad (g_i g_j g_k g_l)^{-1} \prec g_i^{-1} g_l^{-1} (g_j g_k)^{-1}, \quad (g_i g_j g_k g_l)^{-1} \prec (g_i g_l)^{-1} (g_j g_k)^{-1}, \quad \cdots \ , \nn \\
\eea
which demonstrates the general arguments of spurion analysis developed in Section~\ref{sec:sp}.
When all the couplings in Eq.~\eqref{eq:TY_coup} are switched off, none of the couplings in Eq.~\eqref{eq:TY_coup_2} are radiatively generated. Hence, the couplings in Eq.~\eqref{eq:TY_coup} are technically natural.

In the limit when the couplings in Eq.~\eqref{eq:TY_coup} with nontrivial labeling are switched off, the theory is invariant under the lifted $\mathbb{Z}_N\times \mathbb{Z}_2$ symmetry group. We define the surjective map 
\be
\varphi: \quad \mathbb{Z}_N \times \mathbb{Z}_2 \longmapsto \text{TY}(\mathbb{Z}_N)\;,
\ee
such that the elements of TY$(\mathbb{Z}_N)$ are obtained as the images of the elements in $\mathbb{Z}_N\times \mathbb{Z}_2$. Explicitly, we define  
\be
\varphi \left[(g_i, \hat{\mathcal{N}})\right] = \mathcal{N} \;, \quad\quad
\varphi \left[(g_i, 1)\right] = g_i\;, 
\ee
where $\hat{\mathcal{N}}$ obeys the group law of a $\mathbb{Z}_2$ group, i.e., $\hat{\mathcal{N}}^2=1$. For the dynamical particles in Eq.~\eqref{eq:Lag_TY}, the particle $\mathcal{N}$ is relabled using the group element $(1, \hat{\mathcal{N}}) \in \mathbb{Z}_N \times \mathbb{Z}_2$, the other particles $g_i$ are relabeled as $(g_i, 1)\in \mathbb{Z}_N \times \mathbb{Z}_2$. Conventional spurion analysis suggests 
\begin{equation}
c(\lambda^{(0)}_{g_i\mathcal{N}^2})= \left(g_i^{-1}, 1\right), \quad 
c(\lambda^{(0)}_{g_i g_j \mathcal{N}^2 })= \left((g_i g_j)^{-1}, 1\right),
\end{equation}
while all the other couplings in Eq.~\eqref{eq:Lag_TY} are relabeled as $(1,1)$. The consistency of the relations of the couplings above is manifest following the group law of $\mathbb{Z}_N \times \mathbb{Z}_2$.

While the two labeling schemes, one based on TY$(\mathbb{Z}_N)$ and the other on the lifted $\mathbb{Z}_N \times \mathbb{Z}_2$ group, are consistent, we emphasize the important distinctions regarding the hierarchical structures of $\mathbb{Z}_N$-violating interactions as implied by their selection rules; see the discussions after Eq.~\eqref{eq:Lag_TY}.

\subsection{Rep$(S_3)$}
\label{subsec:rep_s3}

Finally, let us consider the fusion algebra Rep$(S_3)$, which contains three basis elements: $\mathbbm{1}$, $X$, and $Y$. Their fusion rules are summarized in Table~\ref{tab:rep_s3}. In the $G+n^\prime$ notation, this corresponds to the $\mathbb{Z}_2 + 1$ type near-group fusion algebra, where $\mathbbm{1}$ and $X$ generate an ordinary $\mathbb{Z}_2$ subgroup, and $Y$ is a non-invertible element satisfying the generic near-group fusion rules in Eqs.~\eqref{eq:NG_1} and~\eqref{eq:NG_2}.~\footnote{Intuitively, as suggested by the fusion rules in Table~\ref{tab:rep_s3}, Rep$(S_3)$ may be viewed as a hybrid of the Fibonacci and Ising fusion algebras.}

Notice that, although Rep$(S_3)$ originates from the representation theory of the non-Abelian group $S_3$, there are important conceptual differences in using its basis elements to label particles in our context. In particular, we do not assume that particles labeled by these elements form actual irreducible representations of $S_3$.~\footnote{For instance, the element $Y$ corresponds to the two-dimensional representation of $S_3$. However, in our following example, this does not imply that the particle labeled by $Y$ is a doublet.}

\begin{table}[h]
\centering
\renewcommand{\arraystretch}{1.0}
\setlength{\tabcolsep}{12pt}
\begin{tabular}{c|ccc}
 & $\mathbbm{1}$ & $X$ & $Y$ \\
\hline
$\mathbbm{1}$ & $\mathbbm{1}$ & $X$ & $Y$ \\
$X$ & $X$ & $\mathbbm{1}$ & $Y$ \\
$Y$ & $Y$ & $Y$ & $\mathbbm{1} + X + Y$ \\
\end{tabular}
\caption{Fusion algebra of $\text{Rep}(S_3)$ with the basis elements denoted as $\mathbbm{1}$, $X$, and $Y$.}
\label{tab:rep_s3}
\end{table}

Using the basis elements to denote the corresponding particles that they label, the classical Lagrangian that faithfully realizes the Rep$(S_3)$ can be written as
\bea
\mathcal{L}_{\text{Rep}(S_3)} \supset && \lambda^{(0)}_1 \mathbbm{1}+\lambda^{(0)}_{1^2} \mathbbm{1}^2+\lambda^{(0)}_{X^2} X^2 +\lambda^{(0)}_{Y^2} Y^2 \nn\\
&& + \lambda^{(0)}_{1^3} \mathbbm{1}^3+\lambda^{(0)}_{Y^3} Y^3 + \lambda^{(0)}_{X^2 1} X^2 \mathbbm{1} + \lambda^{(0)}_{Y^2 1} Y^2 \mathbbm{1} + \lambda^{(0)}_{Y^2 X} Y^2 X \nn \\ 
&& + \lambda^{(0)}_{1^4} \mathbbm{1}^4 + \lambda^{(0)}_{X^4} X^4 + \lambda^{(0)}_{Y^4} Y^4 + \lambda^{(0)}_{Y^3 1} Y^3 \mathbbm{1} + \lambda^{(0)}_{Y^3 X} Y^3 X  \nn\\
&& + \lambda^{(0)}_{1^2 X^2} \mathbbm{1}^2 X^2 + \lambda^{(0)}_{1^2 Y^2} \mathbbm{1}^2 Y^2 + \lambda^{(0)}_{X^2 Y^2} X^2 Y^2+\lambda^{(0)}_{Y^2 1 X} Y^2 \mathbbm{1} X \;.
\label{eq:Lag_Rep_S3}
\eea
The allowed interaction terms in $\mathcal{L}_{\text{Rep}(S_3)}$ are similar to those in the models based on the Fib. or the TY$(\mathbb{Z}_2)$ (i.e., Ising) fusion algebras. 
More specifically, Eq.~\eqref{eq:Lag_Rep_S3} has the following features. 
\begin{itemize}
\item At tree level, all terms linear in $Y$ are forbidden, while the terms with any higher powers of $Y$ are allowed. However, the latter can induce terms linear in $Y$ at the one-loop level in radiative corrections. 
\item For the terms without involving $Y$, the elements $\mathbbm{1}$ and $X$ obey the group law of an ordinary $\mathbb{Z}_2$ group at tree level. 
\item In contrast, for the terms involving $Y$, the elements $\mathbbm{1}$ and $X$ do not have to obey the group law of $\mathbb{Z}_2$. In turn, these terms induce new interaction terms among only $\mathbbm{1}$ and $X$ at the one-loop level via $Y$ loops. These radiatively generated terms generally violate the $\mathbb{Z}_2$ group.
\end{itemize}
The hierarchical structures --- between linear and higher-power terms in $Y$, and between the $\mathbb{Z}_2$-violating interactions with or without $Y$ --- are the hallmark of the NISRs from Rep$(S_3)$.

Following our labeling rule~\ref{rule1}, we label the couplings using the basis elements of Rep$(S_3)$,
\be
(\lambda^{(0)}_{Y^3}) = Y, \quad 
(\lambda^{(0)}_{Y^2 X}) = X, \quad 
(\lambda^{(0)}_{Y^3 1}) = Y, \quad 
(\lambda^{(0)}_{Y^3 X}) = Y, \quad 
(\lambda^{(0)}_{Y^2 1 X}) = X \;,
\label{eq:Rep_S3_coup}
\ee
while all the other couplings are labeled by the identity. Although $\mathcal{L}_{\text{Rep}(S_3)}$ fully respects Rep$(S_3)$, the couplings in Eq.~\eqref{eq:Rep_S3_coup} are labeled nontrivially. It implies that radiative corrections involving these couplings can generate amplitudes violating Rep$(S_3)$ at the loop level, and eventually reducing the fusion algebra to identity. Such a result is consistent with~\cite{Kaidi:2024wio}. 
Furthermore, as we briefly mentioned after Eq.~\eqref{eq:cl_Lag}, $\mathcal{L}_{\text{Rep}(S_3)}$ matches $\mathcal{L}_{\text{TY}(\mathbb{Z}_2)}$ once all the interactions with odd powers of $Y$ are turned off. This amounts to turning off the couplings labeled by $Y$ in Eq.~\eqref{eq:Rep_S3_coup}. We remark that this example can be viewed as a nested partial ordering of the NISRs: $\text{Rep}(S_3) \rightarrow \text{TY}(\mathbb{Z}_2)\rightarrow \mathbb{Z}_2 \times \mathbb{Z}_2$.

\begin{figure}[t]
\centering
\includegraphics[scale=0.20]{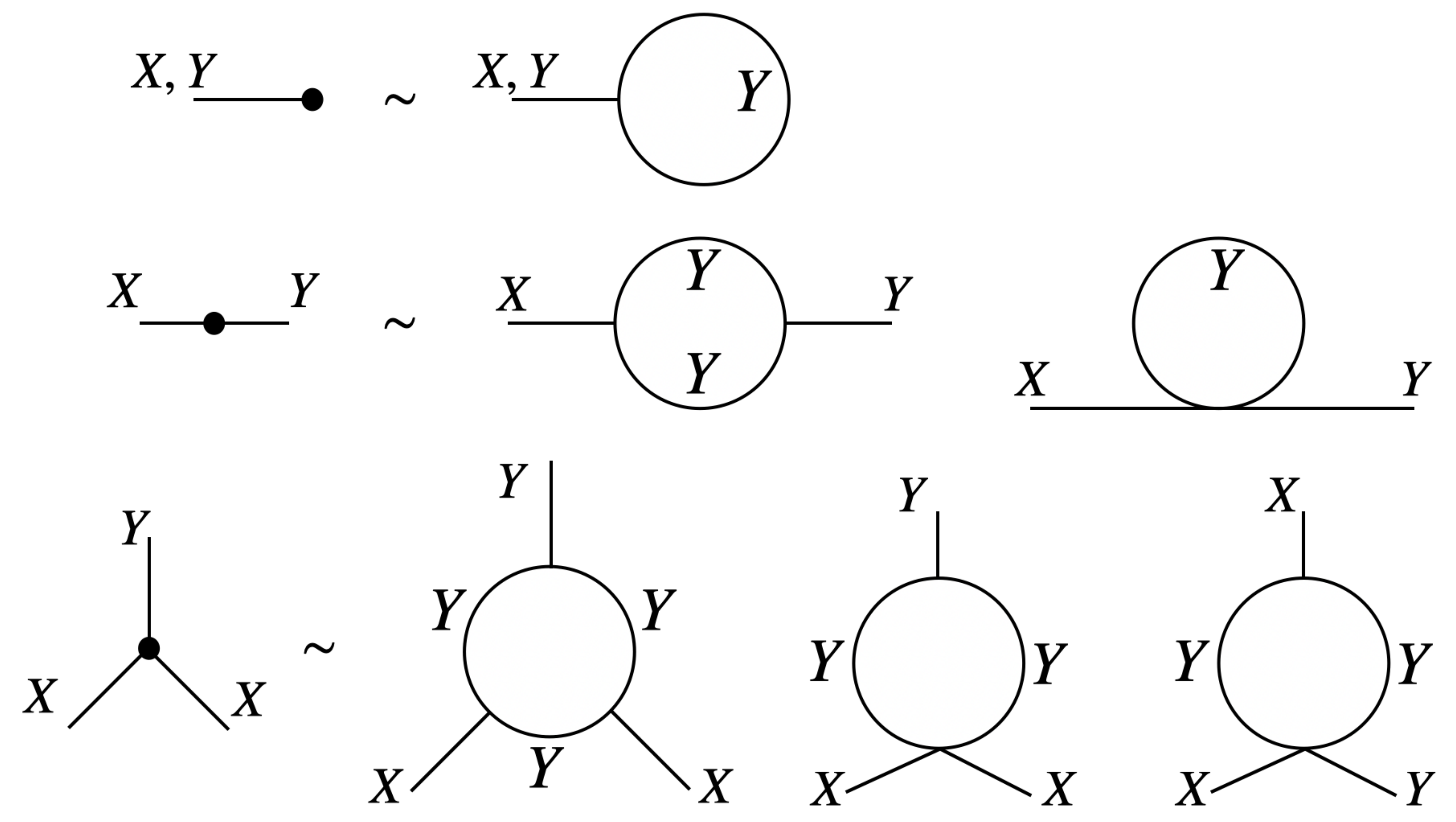}
\caption{A cartoon illustrating some of the one-point, two-point, three-point scattering processes that correspond to the operators forbidden by Rep$(S_3)$ at tree level. 
}
\label{fig:RepS3_1}
\end{figure}

\begin{figure}[t]
\centering
\includegraphics[scale=0.20]{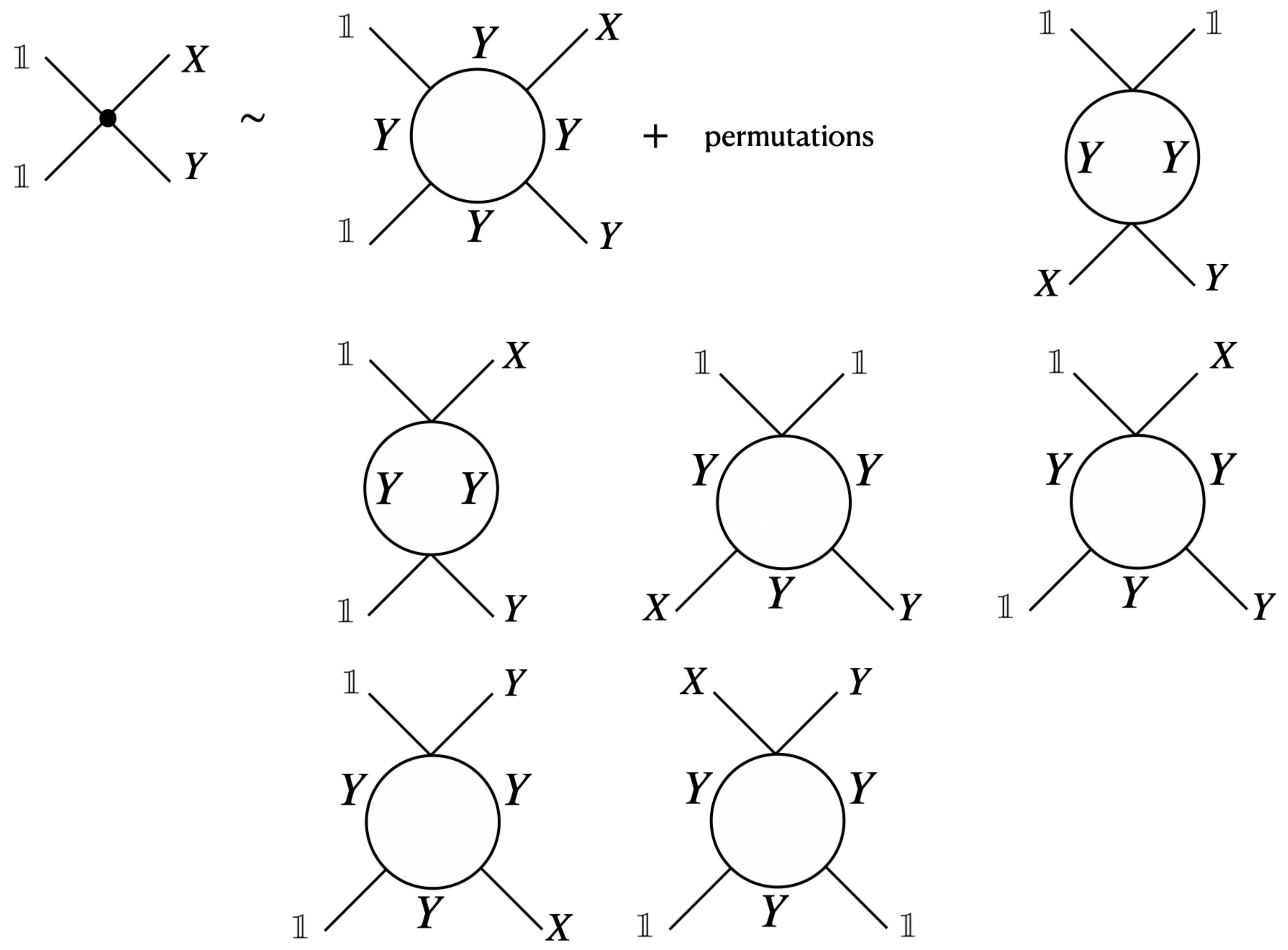}
\caption{A cartoon illustrating the four-point scattering amplitude that corresponds to the operator $\mathbbm{1}^2 X Y$, which is forbidden by Rep$(S_3)$ at tree level.
}
\label{fig:RepS3_2}
\end{figure}

Following the spurion analysis in Section~\ref{sec:sp}, the labeling in Eq.~\eqref{eq:Rep_S3_coup} enables us to systematically track these couplings in all scattering processes. 
For instance, we consider amplitudes associated with the operators linear in $Y$,
\be
Y, X Y, Y \mathbbm{1}, \mathbbm{1}^2 Y,
\mathbbm{1} X Y, X^2 Y, \mathbbm{1}^3 Y, \mathbbm{1}^2 X Y, X^2 \mathbbm{1} Y, X^3 Y 
\ee
and those involving only $\mathbbm{1}$ and $X$ but violating the $\mathbb{Z}_2$ symmetry group,
\be
X, X \mathbbm{1}, X^3, \mathbbm{1}^2 X, \mathbbm{1}^3 X, X^3 \mathbbm{1}, 
\ee
which are all forbidden at tree level but generated at one-loop level through radiative corrections. As shown in Figs.~\ref{fig:RepS3_1} and~\ref{fig:RepS3_2}, we consider the terms $X$, $Y$, $X Y$, $X^2 Y$, and $\mathbbm{1}^2 X Y$ as concrete examples to test the general arguments in Section~\ref{sec:sp}. According to the labeling rule~\ref{rule1}, their couplings are labeled as
\be
c(\lambda^{(1)}_{X}) = X, \quad 
c(\lambda^{(1)}_{Y}) = Y, \quad 
c(\lambda^{(1)}_{X Y}) = Y, \quad
c(\lambda^{(1)}_{X^2 Y}) = Y, \quad
c(\lambda^{(1)}_{1^2 X Y}) =  Y.
\label{eq:Rep_S3_coup_2}
\ee
As shown in Figs.~\ref{fig:RepS3_1} and~\ref{fig:RepS3_2}, the associated amplitudes can be constructed from the tree vertices in Eq.~\eqref{eq:Lag_Rep_S3}, leading to the relations between couplings as
\bea
\lambda^{(1)}_{X} &= & \frac{1}{16\pi^2} \lambda^{(0)}_{X Y^2},\quad \lambda^{(1)}_{Y} =  \frac{1}{16\pi^2} \lambda^{(0)}_{Y^3}, \label{eq:repS3_loop1}\\
\lambda^{(1)}_{X Y} & = & \frac{1}{16\pi^2} \left(\lambda^{(0)}_{Y^2 X} \lambda^{(0)}_{Y^3}+\lambda^{(0)}_{Y^3 X}\right),\label{eq:repS3_loop2}\\
\lambda^{(1)}_{X^2 Y} & = & \frac{1}{16\pi^2} \left(\left(\lambda^{(0)}_{Y^2 X}\right)^2 \lambda^{(0)}_{Y^3} + \lambda^{(0)}_{Y^3} \lambda^{(0)}_{X^2 Y^2} + \lambda^{(0)}_{Y^2 X} \lambda^{(0)}_{Y^3 X} \right),\label{eq:repS3_loop3}
\eea
\bea
\lambda^{(1)}_{1^2 X Y} & = & \frac{1}{16\pi^2} \left( \left[\left(\lambda^{(0)}_{Y^2 1}\right)^2 \lambda^{(0)}_{Y^2 X} \lambda^{(0)}_{Y^3} +\text{permutations} \right] + \lambda^{(0)}_{1^2 Y^2} \lambda^{(0)}_{Y^3 X} + \lambda^{(0)}_{Y^2 1 X} \lambda^{(0)}_{Y^3 1} \right.\nn\\
&& \left. + \lambda^{(0)}_{1^2 Y^2} \lambda^{(0)}_{Y^2 X} \lambda^{(0)}_{Y^3} + \lambda^{(0)}_{Y^2 1 X} \lambda^{(0)}_{Y^2 1} \lambda^{(0)}_{Y^3} + \lambda^{(0)}_{Y^3 1} \lambda^{(0)}_{Y^2 1} \lambda^{(0)}_{Y^2 X} + 
\lambda^{(0)}_{Y^3 X} (\lambda^{(0)}_{Y^2 1} )^2   \right),\label{eq:repS3_loop4}
\eea
which are consistent with the fusion algebra of Rep$(S_3)$, i.e.,
\bea
X &\prec & X, \quad Y \prec  Y, \\
Y &\prec & X Y, \quad Y\prec Y, \\
Y &\prec & X^2 Y , \quad Y\prec Y \mathbbm{1}, \quad Y\prec X Y,\\
Y &\prec & \mathbbm{1}^2 X Y ,\quad Y \prec \mathbbm{1} Y, \quad Y \prec X Y, \quad Y \prec \mathbbm{1} X Y, \quad Y \prec X \mathbbm{1} Y, \quad Y \prec Y \mathbbm{1} X ,\quad Y \prec  Y \mathbbm{1}^2.
\eea
These equations imply that the couplings in Eq.~\eqref{eq:Rep_S3_coup_2} are not generated in radiative corrections if all the couplings in Eq.~\eqref{eq:Rep_S3_coup} are switched off, ensuring the latter are technically natural. 
Beyond the examples discussed above, the general arguments of spurion analysis can be tested through other processes.

In the limit where the couplings in Eq.~\eqref{eq:Rep_S3_coup} are switched off, the theory $\mathcal{L}_{\text{Rep}(S_3)}$ is invariant under the lifted $\mathbb{Z}_2\times \mathbb{Z}_2$ symmetry group, where these two $\mathbb{Z}_2$ groups are generated by $X$ and $\hat Y$, i.e. $X^2={\hat Y}^2=1$. We define the surjective map 
\be
\varphi: \quad  \mathbb{Z}_2\times \mathbb{Z}_2 \longmapsto \text{Rep}(S_3)\;,
\ee
such that the elements of Rep$(S_3)$ are obtained as the images of the elements in $\mathbb{Z}_2\times \mathbb{Z}_2$. Following Eq.~\eqref{eq:map}, we define
\be
\varphi\left[(1,1)\right]=\mathbbm{1}, \quad \varphi\left[(X,1)\right]=X, \quad \varphi\left[(1,\hat{Y})\right]=Y, \quad \varphi\left[(X,\hat{Y})\right]=Y.
\ee
For the dynamical particles in Eq.~\eqref{eq:Lag_Rep_S3}, the particle $\mathbbm{1}$ is relabeled by $(1,1)\in \mathbb{Z}_2\times \mathbb{Z}_2$, the particle $X$ is relabeled by $(X,1)\in \mathbb{Z}_2\times \mathbb{Z}_2$, the particle $Y$ is relabeled by $(1,\hat{Y})\in \mathbb{Z}_2\times \mathbb{Z}_2$, respectively. Accordingly, the couplings are relabeled by the inverse total charge of the dynamical particles, i.e., 
\be
c(\lambda^{(0)}_{Y^3}) = (1,\hat{Y}), \quad c(\lambda^{(0)}_{Y^2 X}) = (X,1), \quad 
c(\lambda^{(0)}_{Y^3 1}) = (1,\hat{Y}), \quad c(\lambda^{(0)}_{Y^3 X}) = (X,\hat{Y}), \quad c(\lambda^{(0)}_{Y^2 1 X}) = (X,1) \;,
\label{eq:Rep_S3_coup_3}
\ee
all the other couplings are $(1,1)$ under $\mathbb{Z}_2\times \mathbb{Z}_2$. 
The consistency of the Eqs.~\eqref{eq:repS3_loop1} -- \eqref{eq:repS3_loop4} is manifest following the group law of $\mathbb{Z}_2\times \mathbb{Z}_2$.

While the two labeling schemes, one based on Rep$(S_3)$ and the other on the lifted $\mathbb{Z}_2\times \mathbb{Z}_2$ group, are mutually consistent, we emphasize the distinctions regarding the hierarchical structures implied by Rep$(S_3)$ that cannot be explained by $\mathbb{Z}_2\times \mathbb{Z}_2$, as discussed after Eq.~\eqref{eq:Lag_Rep_S3}.

\section{Conclusion and outlook}
\label{sec:conc}

In this paper, we generalize the framework of spurion analysis to a class of NISRs from near-group fusion algebras of the type $G+n^\prime$~\cite{Evans:2012ta}. 
\begin{itemize}
\item Unlike the case for Abelian symmetries obeying the group law, here the coupling constants need to be labeled by the nontrivial basis elements even when the non-invertible fusion algebra is exact in the classical Lagrangian. This enables us to perform the spurion analysis without relying on the limit where the nontrivially-labeled couplings are switched off. Based on it, we provide an intuitive interpretation of ``loop-induced groupification'' for the near-group fusion algebras, complementary to the perspective in~\cite{Kaidi:2024wio}. 
\item Furthermore, we identify the lifted $G\times \mathbb{Z}_2$ group, which provides an alternative scheme of spurion analysis consistent with the one based on the $G+n^\prime$ fusion algebra. However, crucial differences remain: the hierarchical structure of the couplings implied by NISRs from the $G+n^\prime$ fusion algebra cannot be explained by the selection rules from the lifted $G\times \mathbb{Z}_2$ group. 
\item Motivated by the particle physics applications such as those in~\cite{Suzuki:2025oov}, we also demonstrate the general framework with a few concrete examples. 
\end{itemize}

There are several directions to extend the results of the current work. 
First, it is natural to generalize the spurion analysis to NISRs beyond near-group fusion algebras. New features are expected when multiple non-invertible elements are present. As a very preliminary step, we present an example based on the fusion algebra of the conjugacy class of $S_3$ (i.e., Conj$(S_3)$) in Appendix~\ref{app_conj_s3}, while a more systematic survey is left to future work~\cite{Suzuki:2025kxz, Xu:2026nwh}.
Secondly, it is interesting to explore the ultraviolet origin of NISRs from an effective field theory perspective. For instance, one may begin with a theory that realizes a fusion algebra faithfully and study how the NISRs are deformed or become unfaithful after integrating out the heavy particles~\cite{Suzuki:2025kxz}. In this way, one may systematically understand the relations between different NISRs along the perturbative renormalization group flows.
In the end, it is important to survey the roles of NISRs in particle phenomenology, see e.g.~\cite{Kobayashi:2024yqq, Kobayashi:2024cvp, Funakoshi:2024uvy, Kobayashi:2025znw, Suzuki:2025oov, Liang:2025dkm, Kobayashi:2025ldi, Kobayashi:2025cwx, Nomura:2025sod, Kobayashi:2025lar, Dong:2025jra, Nomura:2025yoa, Chen:2025awz, Okada:2025kfm, Kobayashi:2025thd} for recent works.

\section{Acknowledgment}
We thank Lian-Tao Wang for helpful discussions and comments. 
We thank Justin Kaidi and Yuji Tachikawa for helpful comments on an earlier version of the draft.
M.S. is supported by the MUR projects 2017L5W2PT. M.S. also acknowledges the European Union - NextGenerationEU, in the framework of the PRIN Project “Charting unexplored avenues in Dark Matter” (20224JR28W).
The work of L.X.X. is partially supported by ``Exotic High Energy Phenomenology" (X-HEP), a project funded by the European Union - Grant Agreement n.101039756. Funded by the European Union. Views and opinions expressed are however those of the author(s) only and do not necessarily reflect those of the European Union or the ERC Executive Agency (ERCEA). Neither the European Union nor the granting authority can be held responsible for them.
This project is also supported by the Munich Institute for Astro-, Particle and BioPhysics (MIAPbP) which is funded by the Deutsche Forschungsgemeinschaft (DFG, German Research Foundation) under Germany's Excellence Strategy – EXC-2094 – 390783311. H.Y.Z. is supported by WPI Initiative, MEXT, Japan at Kavli IPMU, the University of Tokyo.

\begin{appendix}
\section{Conj$(S_3)$ as an example beyond near-group fusion algebras}
\label{app_conj_s3}
In this appendix, we survey the non-invertible fusion algebra where particles are labeled by the conjugacy classes of the non-Abelian group $S_3$, which we denote as Conj$(S_3)$. As in Section~\ref{sec:exp}, we consider only gauge-singlet real scalar particles, such that their interactions are only constrained by the NISRs, but not other selection rules imposed by other quantum numbers.

The fusion algebra of Conj$(S_3)$ consists of three basis elements corresponding to the three conjugacy classes of the group $S_3$: $[1]=\{e\}$, $[a]=\{(123),(132)\}$, $[b]=\{(12), (13), (23)\}$. The fusion rules are summarized in Table~\ref{tab:conj_s3}, where both the basis elements $[a]$ and $[b]$ are non-invertible, while the identity element $[1]$ forms a trivial group. Hence, Conj$(S_3)$ is arguably the simplest example beyond the near-group fusion algebra. 

\begin{table}[h]
\centering
\renewcommand{\arraystretch}{1.0}
\setlength{\tabcolsep}{12pt}
\begin{tabular}{c|ccc}
 & $[1]$ & $[a]$ & $[b]$ \\
\hline
$[1]$ & $[1]$ & $[a]$ & $[b]$ \\
$[a]$ & $[a]$ & $[a]+2[1]$ & $2[b]$ \\
$[b]$ & $[b]$ & $2[b]$ & $3[1]+3[a]$ \\
\end{tabular}
\caption{Fusion algebra of Conj$(S_3)$ with the basis elements denoted as $[1]$, $[a]$, and $[b]$.}
\label{tab:conj_s3}
\end{table}

To faithfully realize the Conj$(S_3)$, we introduce three real scalars which are labeled by $\phi_1\sim [1]$, $\phi_2\sim [a]$, and $\phi_3\sim [b]$, respectively. The classical Lagrangian is then 
\bea
\mathcal{L}_{\text{Conj}(S_3)} \supset && \lambda^{(0)}_{[1]} \phi_1 + \lambda^{(0)}_{[1]^2} \phi_1^2 + \lambda^{(0)}_{[a]^2} \phi_2^2 + \lambda^{(0)}_{[b]^2} \phi_3^2 \nn\\
&& + \lambda^{(0)}_{[1]^3} \phi_1^3 + \lambda^{(0)}_{[a]^3} \phi_2^3 + \lambda^{(0)}_{[a]^2 [1]} \phi_2^2 \phi_1 + \lambda^{(0)}_{[b]^2 [1]} \phi_3^2 \phi_1 + \lambda^{(0)}_{[b]^2[a]} \phi_3^2 \phi_2 \nn\\ 
&& + \lambda^{(0)}_{[1]^4} \phi_1^4+ \lambda^{(0)}_{[a]^4} \phi_2^4 + \lambda^{(0)}_{[b]^4} \phi_3^4 + \lambda^{(0)}_{[a]^3 [1]} \phi_2^3 \phi_1+ \lambda^{(0)}_{[1][a][b]^2} \phi_1\phi_2\phi_3^2 \nn\\
&& + \lambda^{(0)}_{[1]^2[a]^2} \phi_1^2\phi_2^2 +\lambda^{(0)}_{[1]^2 [b]^2} \phi_1^2 \phi_3^2 +\lambda^{(0)}_{[a]^2 [b]^2} \phi_2^2 \phi_3^2\ .
\label{eq:Lag_Conj_S3}
\eea
The allowed interaction terms in  $\mathcal{L}_{\text{Conj}(S_3)}$ have the following features. 
\begin{itemize}

\item Among the terms involving only $\phi_1$ and $\phi_2$, those linear in $\phi_2$ are forbidden at tree level, while those with higher powers are allowed. Indeed, the fusion rules among the elements $[1]$ and $[a]$ are similar to those of the Fib; see in Section~\ref{subsec:fib}. As in the Fib case, $\phi_1$ and $\phi_2$ are identified at the one-loop order, i.e., $\phi_1\prec \phi_2 (\phi_2)^2$, suggesting that the terms linear in $\phi_2$ are generated in radiative corrections at one-loop order.

\item Only the terms with even powers of $\phi_3$ are allowed. This is consistent with~\cite{Kaidi:2024wio}, which indicates an all-order $\mathbb{Z}_2$ group under which $\phi_3\to -\phi_3$.
Furthermore, the terms involving $\phi_3$ resemble those in the TY$(\mathbb{Z}_N)$ case studied in~\ref{subsec:TY}.
Indeed, the fusion rules involving $[b]$ are similar to those of the non-invertible element $\mathcal{N}$ in Eq.~\eqref{eq:TY_ZN_fusion_rules}, although the remaining elements do not form a group.
\end{itemize}
Hence, although Conj$(S_3)$ is not a near-group fusion algebra, it can intuitively be viewed as another hybrid of the Fibonacci and Ising (i.e., TY$(\mathbb{Z}_2)$) fusion algebras.

Inspired by such similarity, we adopt the same labeling rule~\ref{rule1} to label the couplings in Eq.~\eqref{eq:Lag_Conj_S3}, although we do not have a proof that these rules remain consistent when applied in spurion analysis beyond near-group fusion algebras.~\footnote{This needs to be distinguished from the case of near-group fusion algebras, where in Section~\ref{sec:gene_sp} we have demonstrated that the labeling rule~\ref{rule1} is indeed a consistent prescription in spurion analysis.} 
The couplings in Eq.~\eqref{eq:Lag_Conj_S3} are labeled as
\be
c(\lambda^{(0)}_{[a]^3})= [a], \quad c(\lambda^{(0)}_{[b]^2[a]})= [a], \quad 
c(\lambda^{(0)}_{[a]^3[1]})= [a], \quad 
c(\lambda^{(0)}_{[1][a][b]^2})= [a],  
\label{eq:Conj_S3_coup}
\ee
while all the other couplings are labeled by the identity element. Based on Eq.~\eqref{eq:Conj_S3_coup}, we see immediately that the $\mathbb{Z}_2$ group under which $[b]\to -[b]$ is exact up to all orders in perturbation theory: if any amplitude with an odd number of $\phi_3$ particles were generated radiatively, its overall coupling would be labeled by $[b]$. However, the element $[b]$ cannot be obtained from the fusion product of any powers of the couplings in Eq.~\eqref{eq:Conj_S3_coup}.

\begin{figure}[t]
\centering
\includegraphics[scale=0.20]{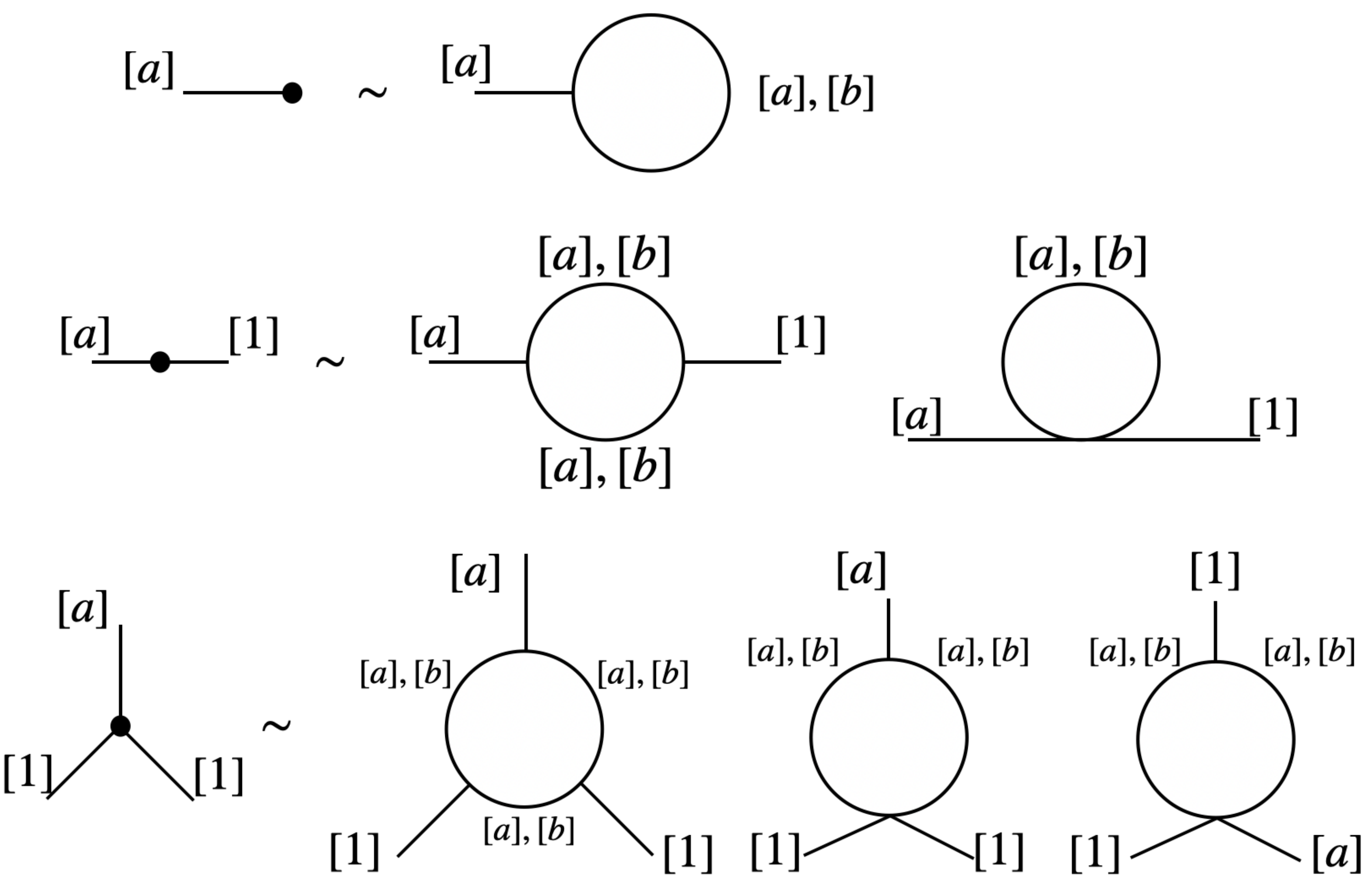}
\caption{A cartoon illustrating some of the one-point, two-point, three-point scattering processes that correspond to the operators forbidden by Conj$(S_3)$ at tree level. 
}
\label{fig:ConjS3_1}
\end{figure}

\begin{figure}[t]
\centering
\includegraphics[scale=0.20]{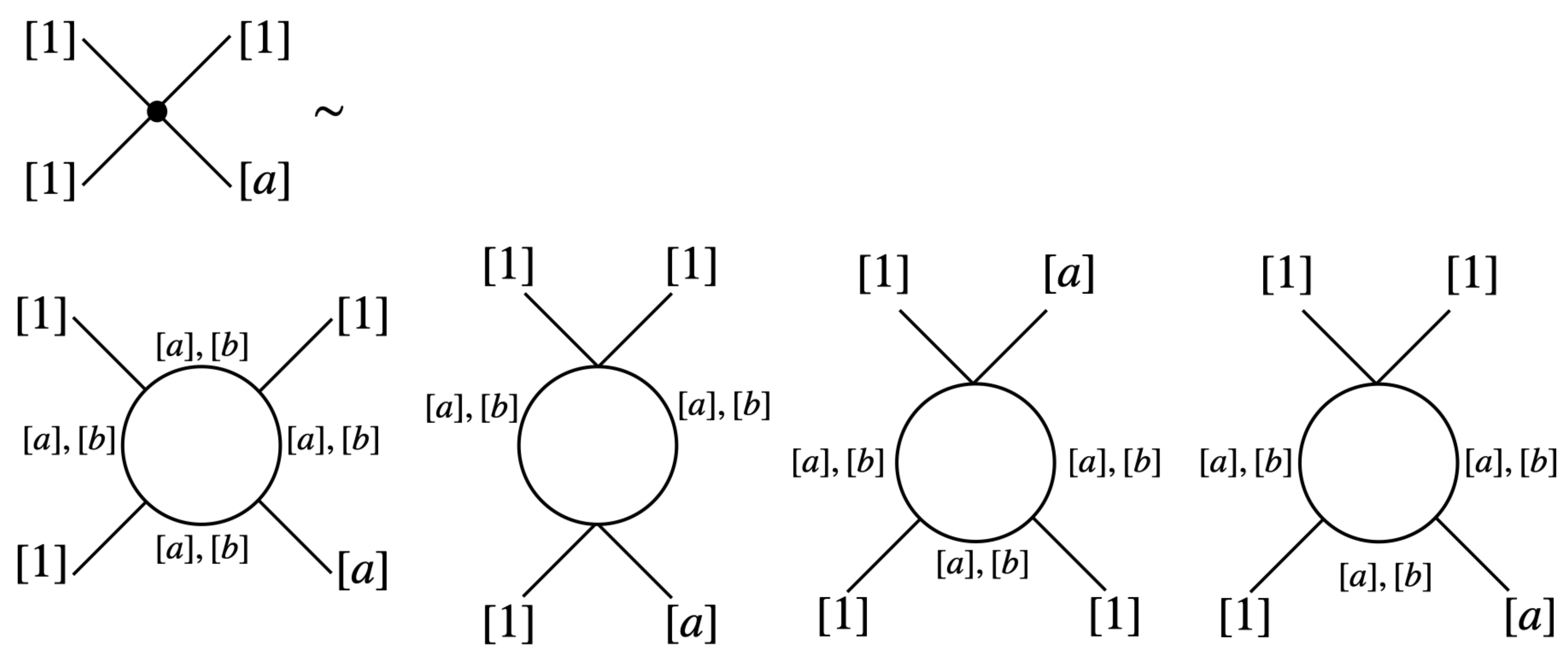}
\caption{A cartoon illustrating the four-point scattering amplitude that corresponds to the operator $\phi_1^3\phi_2$, which is forbidden by Conj$(S_3)$ at tree level.
}
\label{fig:ConjS3_2}
\end{figure}

Next, we consider the one-loop amplitudes corresponding to the operators that are forbidden by Conj$(S_3)$ at tree level but are generated in radiative corrections, in order to test the validity of our labeling rule~\ref{rule1} beyond near-group fusion algebras. The terms that we consider are $\phi_2$, $\phi_1\phi_2$, $\phi_1^2 \phi_2$, and $\phi_1^3 \phi_2$. According to the labeling rule, their couplings are labeled as
\be
c(\lambda^{(1)}_{[a]})= [a], \quad c(\lambda^{(1)}_{[a][1]})= [a], \quad c(\lambda^{(1)}_{[a][1]^2})= [a], \quad c(\lambda^{(1)}_{[a][1]^3})= [a] \;.
\label{eq:Conj_S3_coup_2}
\ee
As shown in Figs.~\ref{fig:ConjS3_1} and~\ref{fig:ConjS3_2}, the associated amplitudes can be constructed from the tree vertices in Eq.~\eqref{eq:Lag_Conj_S3}, leading to the relations between the couplings as
\bea
\lambda^{(1)}_{[a]} = && \frac{1}{16\pi^2} \left(\lambda^{(0)}_{[a]^3}+\lambda^{(0)}_{[b]^2[a]}\right)\;,\\
\lambda^{(1)}_{[a][1]} = && \frac{1}{16\pi^2} \left( \lambda^{(0)}_{[a]^3} \lambda^{(0)}_{[a]^2[1]} + \lambda^{(0)}_{[b]^2[a]} \lambda^{(0)}_{[b]^2[1]} +\lambda^{(0)}_{[1][a][b]^2} + \lambda^{(0)}_{[a]^3[1]} \right)\;, \\
\lambda^{(1)}_{[a][1]^2} = && \frac{1}{16\pi^2} \left( \lambda^{(0)}_{[a]^3} \left(\lambda^{(0)}_{[a]^2 [1]}\right)^2 + \lambda^{(0)}_{[a]^3} \lambda^{(0)}_{[a]^2[1]^2} + \lambda^{(0)}_{[a]^2[1]} \lambda^{(0)}_{[a]^3[1]} \right.\nn\\
\quad && \left. + \lambda^{(0)}_{[a][b]^2} \left(\lambda^{(0)}_{[b]^2 [1]}\right)^2 + \lambda^{(0)}_{[a][b]^2} \lambda^{(0)}_{[b]^2 [1]^2} + \lambda^{(0)}_{[b]^2[1]} \lambda^{(0)}_{[1][a][b]^2} \right)\;, \\
\lambda^{(1)}_{[a][1]^3} = && \frac{1}{16\pi^2} \left( \lambda^{(0)}_{[a]^3} \left(\lambda^{(0)}_{[a]^2 [1]}\right)^3 + \lambda^{(0)}_{[a]^3[1]} \lambda^{(0)}_{[a]^2[1]^2} + \left(\lambda^{(0)}_{[a]^2[1]}\right)^2 \lambda^{(0)}_{[a]^3[1]}+ \lambda^{(0)}_{[a]^2[1]} \lambda^{(0)}_{[a]^2[1]^2} \lambda^{(0)}_{[a]^3} \right.\nn\\
\quad && \left. + \lambda^{(0)}_{[b]^2[a]} \left(\lambda^{(0)}_{[b]^2 [1]}\right)^3 + \lambda^{(0)}_{[b]^2[1][a]} \lambda^{(0)}_{[b]^2[1]^2} + \left(\lambda^{(0)}_{[b]^2[1]}\right)^2 \lambda^{(0)}_{[a][1][b]^2}+ \lambda^{(0)}_{[b]^2[1]} \lambda^{(0)}_{[b]^2[1]^2} \lambda^{(0)}_{[b]^2[a]} \right)\;,\nn\\
\eea
which are consistent with the fusion algebra of Conj$(S_3)$, i.e.,
\begin{align}
[a] &\prec [a], \ [a] \prec [a] \;, \\
[a] &\prec [a][1], \ [a] \prec [a][1], \ [a] \prec [a] , \ [a] \prec [a] \;, \\
[a] &\prec  [a] [1]^2, \ [a] \prec [a] [1], \ [a] \prec [1][a], \ [a] \prec [a] [1]^2, \ [a] \prec [a][1], \ [a] \prec [1][a] \;,\\
[a] &\prec [a] [1]^3, \ [a] \prec [a][1], \ [a] \prec [1]^2 [a], \ [a] \prec [1] [1] [a], \ [a] \prec [a] [1]^3, \ [a] \prec [a] [1], \ [a] \prec [1]^2 [a], \ [a] \prec [1] [1] [a]\ .
\end{align}
These equations imply that the couplings in Eq.~\eqref{eq:Conj_S3_coup_2} are not radiatively generated once all the couplings in Eq.~\eqref{eq:Conj_S3_coup} are switched off, ensuring the latter are technically natural. 

In the limit when the couplings in Eq.~\eqref{eq:Conj_S3_coup} with nontrivial labeling are switched off, the theory is invariant under the lifted $\mathbb{Z}_2\times \mathbb{Z}_2$ symmetry group. We define the surjective map 
\be
\varphi: \quad \mathbb{Z}_2 \times \mathbb{Z}_2 \longmapsto \text{Conj}(S_3)\;,
\ee
such that the elements of Conj$(S_3)$ are obtained as the images of the elements in $\mathbb{Z}_2\times \mathbb{Z}_2$. Explicitly, we define  
\be
\varphi \left[(1, 1)\right] = [1] \;, \quad\quad
\varphi \left[(\hat{a}, 1)\right] = [a]\;, \quad\quad
\varphi \left[(1, \hat{b})\right] = [b] \; \quad\quad
\varphi \left[(\hat{a}, \hat{b})\right] = [b] \;,
\ee
where $\hat{a}$ and $\hat{b}$ obey the group law of a $\mathbb{Z}_2$ group, i.e., $\hat{a}^2=\hat{b}^2=1$. For the dynamical particles in Eq.~\eqref{eq:Lag_Conj_S3}, $\phi_1 \sim (1,1)\in \mathbb{Z}_2\times \mathbb{Z}_2$, $\phi_2 \sim (\hat{a},1)\in \mathbb{Z}_2\times \mathbb{Z}_2$, and $\phi_3 \sim (1, \hat{b})\in \mathbb{Z}_2\times \mathbb{Z}_2 $. Conventional spurion analysis suggests 
\begin{equation}
c(\lambda^{(1)}_{[a]})= (\hat{a},1), \quad c(\lambda^{(1)}_{[a][1]})= (\hat{a},1), \quad c(\lambda^{(1)}_{[a][1]^2})= (\hat{a},1), \quad c(\lambda^{(1)}_{[a][1]^3})= (\hat{a},1) ,
\end{equation}
while all the other couplings in Eq.~\eqref{eq:Lag_Conj_S3} are relabeled as $(1,1)$. The consistency of the relations of the couplings above is also manifest following the group law of $\mathbb{Z}_2 \times \mathbb{Z}_2$.

We emphasize that, while the two schemes of spurion analysis, one based on Conj$(S_3)$ and the other on $\mathbb{Z}_2 \times \mathbb{Z}_2$ are consistent, the hierarchical structures of the couplings implied by the NISRs from Conj$(S_3)$ cannot be explained by breaking the $\mathbb{Z}_2 \times \mathbb{Z}_2$ group. This feature is similar to both the Fibonacci and Ising fusion algebras.

\end{appendix}

\bibliography{NoninvertSpurion.bib}

\begin{thebibliography}{44}%
\makeatletter
\providecommand \@ifxundefined [1]{%
 \@ifx{#1\undefined}
}%
\providecommand \@ifnum [1]{%
 \ifnum #1\expandafter \@firstoftwo
 \else \expandafter \@secondoftwo
 \fi
}%
\providecommand \@ifx [1]{%
 \ifx #1\expandafter \@firstoftwo
 \else \expandafter \@secondoftwo
 \fi
}%
\providecommand \natexlab [1]{#1}%
\providecommand \enquote  [1]{``#1''}%
\providecommand \bibnamefont  [1]{#1}%
\providecommand \bibfnamefont [1]{#1}%
\providecommand \citenamefont [1]{#1}%
\providecommand \href@noop [0]{\@secondoftwo}%
\providecommand \href [0]{\begingroup \@sanitize@url \@href}%
\providecommand \@href[1]{\@@startlink{#1}\@@href}%
\providecommand \@@href[1]{\endgroup#1\@@endlink}%
\providecommand \@sanitize@url [0]{\catcode `\\12\catcode `\$12\catcode
  `\&12\catcode `\#12\catcode `\^12\catcode `\_12\catcode `\%12\relax}%
\providecommand \@@startlink[1]{}%
\providecommand \@@endlink[0]{}%
\providecommand \url  [0]{\begingroup\@sanitize@url \@url }%
\providecommand \@url [1]{\endgroup\@href {#1}{\urlprefix }}%
\providecommand \urlprefix  [0]{URL }%
\providecommand \Eprint [0]{\href }%
\providecommand \doibase [0]{http://dx.doi.org/}%
\providecommand \selectlanguage [0]{\@gobble}%
\providecommand \bibinfo  [0]{\@secondoftwo}%
\providecommand \bibfield  [0]{\@secondoftwo}%
\providecommand \translation [1]{[#1]}%
\providecommand \BibitemOpen [0]{}%
\providecommand \bibitemStop [0]{}%
\providecommand \bibitemNoStop [0]{.\EOS\space}%
\providecommand \EOS [0]{\spacefactor3000\relax}%
\providecommand \BibitemShut  [1]{\csname bibitem#1\endcsname}%
\let\auto@bib@innerbib\@empty
\bibitem [{\citenamefont {Gaiotto}\ \emph {et~al.}(2015)\citenamefont
  {Gaiotto}, \citenamefont {Kapustin}, \citenamefont {Seiberg},\ and\
  \citenamefont {Willett}}]{Gaiotto:2014kfa}%
  \BibitemOpen
  \bibfield  {author} {\bibinfo {author} {\bibfnamefont {Davide}\ \bibnamefont
  {Gaiotto}}, \bibinfo {author} {\bibfnamefont {Anton}\ \bibnamefont
  {Kapustin}}, \bibinfo {author} {\bibfnamefont {Nathan}\ \bibnamefont
  {Seiberg}}, \ and\ \bibinfo {author} {\bibfnamefont {Brian}\ \bibnamefont
  {Willett}},\ }\bibfield  {title} {\enquote {\bibinfo {title} {{Generalized
  Global Symmetries}},}\ }\href {\doibase 10.1007/JHEP02(2015)172} {\bibfield
  {journal} {\bibinfo  {journal} {JHEP}\ }\textbf {\bibinfo {volume} {02}},\
  \bibinfo {pages} {172} (\bibinfo {year} {2015})},\ \Eprint
  {http://arxiv.org/abs/1412.5148} {arXiv:1412.5148 [hep-th]} \BibitemShut
  {NoStop}%
\bibitem [{\citenamefont {Cordova}\ \emph {et~al.}(2022)\citenamefont
  {Cordova}, \citenamefont {Dumitrescu}, \citenamefont {Intriligator},\ and\
  \citenamefont {Shao}}]{Cordova:2022ruw}%
  \BibitemOpen
  \bibfield  {author} {\bibinfo {author} {\bibfnamefont {Clay}\ \bibnamefont
  {Cordova}}, \bibinfo {author} {\bibfnamefont {Thomas~T.}\ \bibnamefont
  {Dumitrescu}}, \bibinfo {author} {\bibfnamefont {Kenneth}\ \bibnamefont
  {Intriligator}}, \ and\ \bibinfo {author} {\bibfnamefont {Shu-Heng}\
  \bibnamefont {Shao}},\ }\bibfield  {title} {\enquote {\bibinfo {title}
  {{Snowmass White Paper: Generalized Symmetries in Quantum Field Theory and
  Beyond}},}\ }in\ \href@noop {} {\emph {\bibinfo {booktitle} {{Snowmass
  2021}}}}\ (\bibinfo {year} {2022})\ \Eprint {http://arxiv.org/abs/2205.09545}
  {arXiv:2205.09545 [hep-th]} \BibitemShut {NoStop}%
\bibitem [{\citenamefont {McGreevy}(2023)}]{McGreevy:2022oyu}%
  \BibitemOpen
  \bibfield  {author} {\bibinfo {author} {\bibfnamefont {John}\ \bibnamefont
  {McGreevy}},\ }\bibfield  {title} {\enquote {\bibinfo {title} {{Generalized
  Symmetries in Condensed Matter}},}\ }\href {\doibase
  10.1146/annurev-conmatphys-040721-021029} {\bibfield  {journal} {\bibinfo
  {journal} {Ann. Rev. Condensed Matter Phys.}\ }\textbf {\bibinfo {volume}
  {14}},\ \bibinfo {pages} {57--82} (\bibinfo {year} {2023})},\ \Eprint
  {http://arxiv.org/abs/2204.03045} {arXiv:2204.03045 [cond-mat.str-el]}
  \BibitemShut {NoStop}%
\bibitem [{\citenamefont {Gomes}(2023)}]{Gomes:2023ahz}%
  \BibitemOpen
  \bibfield  {author} {\bibinfo {author} {\bibfnamefont {Pedro R.~S.}\
  \bibnamefont {Gomes}},\ }\bibfield  {title} {\enquote {\bibinfo {title} {{An
  introduction to higher-form symmetries}},}\ }\href {\doibase
  10.21468/SciPostPhysLectNotes.74} {\bibfield  {journal} {\bibinfo  {journal}
  {SciPost Phys. Lect. Notes}\ }\textbf {\bibinfo {volume} {74}},\ \bibinfo
  {pages} {1} (\bibinfo {year} {2023})},\ \Eprint
  {http://arxiv.org/abs/2303.01817} {arXiv:2303.01817 [hep-th]} \BibitemShut
  {NoStop}%
\bibitem [{\citenamefont {Schafer-Nameki}(2024)}]{Schafer-Nameki:2023jdn}%
  \BibitemOpen
  \bibfield  {author} {\bibinfo {author} {\bibfnamefont {Sakura}\ \bibnamefont
  {Schafer-Nameki}},\ }\bibfield  {title} {\enquote {\bibinfo {title} {{ICTP
  lectures on (non-)invertible generalized symmetries}},}\ }\href {\doibase
  10.1016/j.physrep.2024.01.007} {\bibfield  {journal} {\bibinfo  {journal}
  {Phys. Rept.}\ }\textbf {\bibinfo {volume} {1063}},\ \bibinfo {pages} {1--55}
  (\bibinfo {year} {2024})},\ \Eprint {http://arxiv.org/abs/2305.18296}
  {arXiv:2305.18296 [hep-th]} \BibitemShut {NoStop}%
\bibitem [{\citenamefont {Brennan}\ and\ \citenamefont
  {Hong}(2023)}]{Brennan:2023mmt}%
  \BibitemOpen
  \bibfield  {author} {\bibinfo {author} {\bibfnamefont {T.~Daniel}\
  \bibnamefont {Brennan}}\ and\ \bibinfo {author} {\bibfnamefont {Sungwoo}\
  \bibnamefont {Hong}},\ }\bibfield  {title} {\enquote {\bibinfo {title}
  {{Introduction to Generalized Global Symmetries in QFT and Particle
  Physics}},}\ }\href@noop {} {\  (\bibinfo {year} {2023})},\ \Eprint
  {http://arxiv.org/abs/2306.00912} {arXiv:2306.00912 [hep-ph]} \BibitemShut
  {NoStop}%
\bibitem [{\citenamefont {Luo}\ \emph {et~al.}(2024)\citenamefont {Luo},
  \citenamefont {Wang},\ and\ \citenamefont {Wang}}]{Luo:2023ive}%
  \BibitemOpen
  \bibfield  {author} {\bibinfo {author} {\bibfnamefont {Ran}\ \bibnamefont
  {Luo}}, \bibinfo {author} {\bibfnamefont {Qing-Rui}\ \bibnamefont {Wang}}, \
  and\ \bibinfo {author} {\bibfnamefont {Yi-Nan}\ \bibnamefont {Wang}},\
  }\bibfield  {title} {\enquote {\bibinfo {title} {{Lecture notes on
  generalized symmetries and applications}},}\ }\href {\doibase
  10.1016/j.physrep.2024.02.002} {\bibfield  {journal} {\bibinfo  {journal}
  {Phys. Rept.}\ }\textbf {\bibinfo {volume} {1065}},\ \bibinfo {pages} {1--43}
  (\bibinfo {year} {2024})},\ \Eprint {http://arxiv.org/abs/2307.09215}
  {arXiv:2307.09215 [hep-th]} \BibitemShut {NoStop}%
\bibitem [{\citenamefont {Shao}(2023)}]{Shao:2023gho}%
  \BibitemOpen
  \bibfield  {author} {\bibinfo {author} {\bibfnamefont {Shu-Heng}\
  \bibnamefont {Shao}},\ }\bibfield  {title} {\enquote {\bibinfo {title}
  {{What's Done Cannot Be Undone: TASI Lectures on Non-Invertible
  Symmetries}},}\ }\href@noop {} {\  (\bibinfo {year} {2023})},\ \Eprint
  {http://arxiv.org/abs/2308.00747} {arXiv:2308.00747 [hep-th]} \BibitemShut
  {NoStop}%
\bibitem [{\citenamefont {Costa}\ \emph {et~al.}(2024)\citenamefont {Costa}
  \emph {et~al.}}]{Costa:2024wks}%
  \BibitemOpen
  \bibfield  {author} {\bibinfo {author} {\bibfnamefont {Davi}\ \bibnamefont
  {Costa}} \emph {et~al.},\ }\bibfield  {title} {\enquote {\bibinfo {title}
  {{Simons Lectures on Categorical Symmetries}},}\ \ }(\bibinfo {year} {2024})\
  \Eprint {http://arxiv.org/abs/2411.09082} {arXiv:2411.09082 [math-ph]}
  \BibitemShut {NoStop}%
\bibitem [{\citenamefont {Iqbal}(2024)}]{Iqbal:2024pee}%
  \BibitemOpen
  \bibfield  {author} {\bibinfo {author} {\bibfnamefont {Nabil}\ \bibnamefont
  {Iqbal}},\ }\bibfield  {title} {\enquote {\bibinfo {title} {{Jena lectures on
  generalized global symmetries: principles and applications}},}\ \ }(\bibinfo
  {year} {2024})\ \Eprint {http://arxiv.org/abs/2407.20815} {arXiv:2407.20815
  [hep-th]} \BibitemShut {NoStop}%
\bibitem [{\citenamefont {Davighi}(2025)}]{Davighi:2025iyk}%
  \BibitemOpen
  \bibfield  {author} {\bibinfo {author} {\bibfnamefont {Joe}\ \bibnamefont
  {Davighi}},\ }\bibfield  {title} {\enquote {\bibinfo {title} {{Generalised
  Symmetries in Particle Physics}},}\ }\href {\doibase 10.22323/1.481.0076}
  {\bibfield  {journal} {\bibinfo  {journal} {PoS}\ }\textbf {\bibinfo {volume}
  {DISCRETE2024}},\ \bibinfo {pages} {076} (\bibinfo {year} {2025})},\ \Eprint
  {http://arxiv.org/abs/2504.05960} {arXiv:2504.05960 [hep-ph]} \BibitemShut
  {NoStop}%
\bibitem [{\citenamefont {Bhardwaj}\ and\ \citenamefont
  {Tachikawa}(2018)}]{Bhardwaj:2017xup}%
  \BibitemOpen
  \bibfield  {author} {\bibinfo {author} {\bibfnamefont {Lakshya}\ \bibnamefont
  {Bhardwaj}}\ and\ \bibinfo {author} {\bibfnamefont {Yuji}\ \bibnamefont
  {Tachikawa}},\ }\bibfield  {title} {\enquote {\bibinfo {title} {{On finite
  symmetries and their gauging in two dimensions}},}\ }\href {\doibase
  10.1007/JHEP03(2018)189} {\bibfield  {journal} {\bibinfo  {journal} {JHEP}\
  }\textbf {\bibinfo {volume} {03}},\ \bibinfo {pages} {189} (\bibinfo {year}
  {2018})},\ \Eprint {http://arxiv.org/abs/1704.02330} {arXiv:1704.02330
  [hep-th]} \BibitemShut {NoStop}%
\bibitem [{\citenamefont {Chang}\ \emph {et~al.}(2019)\citenamefont {Chang},
  \citenamefont {Lin}, \citenamefont {Shao}, \citenamefont {Wang},\ and\
  \citenamefont {Yin}}]{Chang:2018iay}%
  \BibitemOpen
  \bibfield  {author} {\bibinfo {author} {\bibfnamefont {Chi-Ming}\
  \bibnamefont {Chang}}, \bibinfo {author} {\bibfnamefont {Ying-Hsuan}\
  \bibnamefont {Lin}}, \bibinfo {author} {\bibfnamefont {Shu-Heng}\
  \bibnamefont {Shao}}, \bibinfo {author} {\bibfnamefont {Yifan}\ \bibnamefont
  {Wang}}, \ and\ \bibinfo {author} {\bibfnamefont {Xi}~\bibnamefont {Yin}},\
  }\bibfield  {title} {\enquote {\bibinfo {title} {{Topological Defect Lines
  and Renormalization Group Flows in Two Dimensions}},}\ }\href {\doibase
  10.1007/JHEP01(2019)026} {\bibfield  {journal} {\bibinfo  {journal} {JHEP}\
  }\textbf {\bibinfo {volume} {01}},\ \bibinfo {pages} {026} (\bibinfo {year}
  {2019})},\ \Eprint {http://arxiv.org/abs/1802.04445} {arXiv:1802.04445
  [hep-th]} \BibitemShut {NoStop}%
\bibitem [{\citenamefont {Hamidi}\ and\ \citenamefont
  {Vafa}(1987)}]{Hamidi:1986vh}%
  \BibitemOpen
  \bibfield  {author} {\bibinfo {author} {\bibfnamefont {Shahram}\ \bibnamefont
  {Hamidi}}\ and\ \bibinfo {author} {\bibfnamefont {Cumrun}\ \bibnamefont
  {Vafa}},\ }\bibfield  {title} {\enquote {\bibinfo {title} {{Interactions on
  Orbifolds}},}\ }\href {\doibase 10.1016/0550-3213(87)90006-X} {\bibfield
  {journal} {\bibinfo  {journal} {Nucl. Phys. B}\ }\textbf {\bibinfo {volume}
  {279}},\ \bibinfo {pages} {465--513} (\bibinfo {year} {1987})}\BibitemShut
  {NoStop}%
\bibitem [{\citenamefont {Font}\ \emph {et~al.}(1988)\citenamefont {Font},
  \citenamefont {Ibanez}, \citenamefont {Nilles},\ and\ \citenamefont
  {Quevedo}}]{Font:1988nc}%
  \BibitemOpen
  \bibfield  {author} {\bibinfo {author} {\bibfnamefont {A.}~\bibnamefont
  {Font}}, \bibinfo {author} {\bibfnamefont {Luis~E.}\ \bibnamefont {Ibanez}},
  \bibinfo {author} {\bibfnamefont {Hans~Peter}\ \bibnamefont {Nilles}}, \ and\
  \bibinfo {author} {\bibfnamefont {F.}~\bibnamefont {Quevedo}},\ }\bibfield
  {title} {\enquote {\bibinfo {title} {{On the Concept of Naturalness in String
  Theories}},}\ }\href {\doibase 10.1016/0370-2693(88)91760-1} {\bibfield
  {journal} {\bibinfo  {journal} {Phys. Lett. B}\ }\textbf {\bibinfo {volume}
  {213}},\ \bibinfo {pages} {274--278} (\bibinfo {year} {1988})}\BibitemShut
  {NoStop}%
\bibitem [{\citenamefont {Kobayashi}(1995)}]{Kobayashi:1995py}%
  \BibitemOpen
  \bibfield  {author} {\bibinfo {author} {\bibfnamefont {Tatsuo}\ \bibnamefont
  {Kobayashi}},\ }\bibfield  {title} {\enquote {\bibinfo {title} {{Selection
  rules for nonrenormalizable couplings in superstring theories}},}\ }\href
  {\doibase 10.1016/0370-2693(95)00643-Y} {\bibfield  {journal} {\bibinfo
  {journal} {Phys. Lett. B}\ }\textbf {\bibinfo {volume} {354}},\ \bibinfo
  {pages} {264--270} (\bibinfo {year} {1995})},\ \Eprint
  {http://arxiv.org/abs/hep-ph/9504371} {arXiv:hep-ph/9504371} \BibitemShut
  {NoStop}%
\bibitem [{\citenamefont {Kobayashi}\ \emph {et~al.}(2012)\citenamefont
  {Kobayashi}, \citenamefont {Parameswaran}, \citenamefont {Ramos-Sanchez},\
  and\ \citenamefont {Zavala}}]{Kobayashi:2011cw}%
  \BibitemOpen
  \bibfield  {author} {\bibinfo {author} {\bibfnamefont {Tatsuo}\ \bibnamefont
  {Kobayashi}}, \bibinfo {author} {\bibfnamefont {Susha~L.}\ \bibnamefont
  {Parameswaran}}, \bibinfo {author} {\bibfnamefont {Saul}\ \bibnamefont
  {Ramos-Sanchez}}, \ and\ \bibinfo {author} {\bibfnamefont {Ivonne}\
  \bibnamefont {Zavala}},\ }\bibfield  {title} {\enquote {\bibinfo {title}
  {{Revisiting Coupling Selection Rules in Heterotic Orbifold Models}},}\
  }\href {\doibase 10.1007/JHEP12(2012)049} {\bibfield  {journal} {\bibinfo
  {journal} {JHEP}\ }\textbf {\bibinfo {volume} {05}},\ \bibinfo {pages} {008}
  (\bibinfo {year} {2012})},\ \bibinfo {note} {[Erratum: JHEP 12, 049
  (2012)]},\ \Eprint {http://arxiv.org/abs/1107.2137} {arXiv:1107.2137
  [hep-th]} \BibitemShut {NoStop}%
\bibitem [{\citenamefont {Kaidi}\ \emph {et~al.}(2024)\citenamefont {Kaidi},
  \citenamefont {Tachikawa},\ and\ \citenamefont {Zhang}}]{Kaidi:2024wio}%
  \BibitemOpen
  \bibfield  {author} {\bibinfo {author} {\bibfnamefont {Justin}\ \bibnamefont
  {Kaidi}}, \bibinfo {author} {\bibfnamefont {Yuji}\ \bibnamefont {Tachikawa}},
  \ and\ \bibinfo {author} {\bibfnamefont {Hao~Y.}\ \bibnamefont {Zhang}},\
  }\bibfield  {title} {\enquote {\bibinfo {title} {{On a class of selection
  rules without group actions in field theory and string theory}},}\ }\href
  {\doibase 10.21468/SciPostPhys.17.6.169} {\bibfield  {journal} {\bibinfo
  {journal} {SciPost Phys.}\ }\textbf {\bibinfo {volume} {17}},\ \bibinfo
  {pages} {169} (\bibinfo {year} {2024})},\ \Eprint
  {http://arxiv.org/abs/2402.00105} {arXiv:2402.00105 [hep-th]} \BibitemShut
  {NoStop}%
\bibitem [{\citenamefont {Heckman}\ \emph {et~al.}(2024)\citenamefont
  {Heckman}, \citenamefont {McNamara}, \citenamefont {Montero}, \citenamefont
  {Sharon}, \citenamefont {Vafa},\ and\ \citenamefont
  {Valenzuela}}]{Heckman:2024obe}%
  \BibitemOpen
  \bibfield  {author} {\bibinfo {author} {\bibfnamefont {Jonathan~J.}\
  \bibnamefont {Heckman}}, \bibinfo {author} {\bibfnamefont {Jacob}\
  \bibnamefont {McNamara}}, \bibinfo {author} {\bibfnamefont {Miguel}\
  \bibnamefont {Montero}}, \bibinfo {author} {\bibfnamefont {Adar}\
  \bibnamefont {Sharon}}, \bibinfo {author} {\bibfnamefont {Cumrun}\
  \bibnamefont {Vafa}}, \ and\ \bibinfo {author} {\bibfnamefont {Irene}\
  \bibnamefont {Valenzuela}},\ }\bibfield  {title} {\enquote {\bibinfo {title}
  {{Fate of stringy noninvertible symmetries}},}\ }\href {\doibase
  10.1103/PhysRevD.110.106001} {\bibfield  {journal} {\bibinfo  {journal}
  {Phys. Rev. D}\ }\textbf {\bibinfo {volume} {110}},\ \bibinfo {pages}
  {106001} (\bibinfo {year} {2024})},\ \Eprint
  {http://arxiv.org/abs/2402.00118} {arXiv:2402.00118 [hep-th]} \BibitemShut
  {NoStop}%
\bibitem [{\citenamefont {Kobayashi}\ and\ \citenamefont
  {Otsuka}(2024)}]{Kobayashi:2024yqq}%
  \BibitemOpen
  \bibfield  {author} {\bibinfo {author} {\bibfnamefont {Tatsuo}\ \bibnamefont
  {Kobayashi}}\ and\ \bibinfo {author} {\bibfnamefont {Hajime}\ \bibnamefont
  {Otsuka}},\ }\bibfield  {title} {\enquote {\bibinfo {title} {{Non-invertible
  flavor symmetries in magnetized extra dimensions}},}\ }\href {\doibase
  10.1007/JHEP11(2024)120} {\bibfield  {journal} {\bibinfo  {journal} {JHEP}\
  }\textbf {\bibinfo {volume} {11}},\ \bibinfo {pages} {120} (\bibinfo {year}
  {2024})},\ \Eprint {http://arxiv.org/abs/2408.13984} {arXiv:2408.13984
  [hep-th]} \BibitemShut {NoStop}%
\bibitem [{\citenamefont {Kobayashi}\ \emph {et~al.}(2024)\citenamefont
  {Kobayashi}, \citenamefont {Otsuka},\ and\ \citenamefont
  {Tanimoto}}]{Kobayashi:2024cvp}%
  \BibitemOpen
  \bibfield  {author} {\bibinfo {author} {\bibfnamefont {Tatsuo}\ \bibnamefont
  {Kobayashi}}, \bibinfo {author} {\bibfnamefont {Hajime}\ \bibnamefont
  {Otsuka}}, \ and\ \bibinfo {author} {\bibfnamefont {Morimitsu}\ \bibnamefont
  {Tanimoto}},\ }\bibfield  {title} {\enquote {\bibinfo {title} {{Yukawa
  textures from non-invertible symmetries}},}\ }\href {\doibase
  10.1007/JHEP12(2024)117} {\bibfield  {journal} {\bibinfo  {journal} {JHEP}\
  }\textbf {\bibinfo {volume} {12}},\ \bibinfo {pages} {117} (\bibinfo {year}
  {2024})},\ \Eprint {http://arxiv.org/abs/2409.05270} {arXiv:2409.05270
  [hep-ph]} \BibitemShut {NoStop}%
\bibitem [{\citenamefont {Funakoshi}\ \emph {et~al.}(2024)\citenamefont
  {Funakoshi}, \citenamefont {Kobayashi},\ and\ \citenamefont
  {Otsuka}}]{Funakoshi:2024uvy}%
  \BibitemOpen
  \bibfield  {author} {\bibinfo {author} {\bibfnamefont {Shuta}\ \bibnamefont
  {Funakoshi}}, \bibinfo {author} {\bibfnamefont {Tatsuo}\ \bibnamefont
  {Kobayashi}}, \ and\ \bibinfo {author} {\bibfnamefont {Hajime}\ \bibnamefont
  {Otsuka}},\ }\bibfield  {title} {\enquote {\bibinfo {title} {{Quantum aspects
  of non-invertible flavor symmetries in intersecting/magnetized D-brane
  models}},}\ }\href@noop {} {\  (\bibinfo {year} {2024})},\ \Eprint
  {http://arxiv.org/abs/2412.12524} {arXiv:2412.12524 [hep-th]} \BibitemShut
  {NoStop}%
\bibitem [{\citenamefont {Kobayashi}\ \emph
  {et~al.}(2025{\natexlab{a}})\citenamefont {Kobayashi}, \citenamefont
  {Nishioka}, \citenamefont {Otsuka},\ and\ \citenamefont
  {Tanimoto}}]{Kobayashi:2025znw}%
  \BibitemOpen
  \bibfield  {author} {\bibinfo {author} {\bibfnamefont {Tatsuo}\ \bibnamefont
  {Kobayashi}}, \bibinfo {author} {\bibfnamefont {Yume}\ \bibnamefont
  {Nishioka}}, \bibinfo {author} {\bibfnamefont {Hajime}\ \bibnamefont
  {Otsuka}}, \ and\ \bibinfo {author} {\bibfnamefont {Morimitsu}\ \bibnamefont
  {Tanimoto}},\ }\bibfield  {title} {\enquote {\bibinfo {title} {{More about
  quark Yukawa textures from selection rules without group actions}},}\
  }\href@noop {} {\  (\bibinfo {year} {2025}{\natexlab{a}})},\ \Eprint
  {http://arxiv.org/abs/2503.09966} {arXiv:2503.09966 [hep-ph]} \BibitemShut
  {NoStop}%
\bibitem [{\citenamefont {Liang}\ and\ \citenamefont
  {Yanagida}(2025)}]{Liang:2025dkm}%
  \BibitemOpen
  \bibfield  {author} {\bibinfo {author} {\bibfnamefont {Qiuyue}\ \bibnamefont
  {Liang}}\ and\ \bibinfo {author} {\bibfnamefont {Tsutomu~T.}\ \bibnamefont
  {Yanagida}},\ }\bibfield  {title} {\enquote {\bibinfo {title}
  {{Non-invertible symmetry as an axion-less solution to the strong CP
  problem}},}\ }\href@noop {} {\  (\bibinfo {year} {2025})},\ \Eprint
  {http://arxiv.org/abs/2505.05142} {arXiv:2505.05142 [hep-ph]} \BibitemShut
  {NoStop}%
\bibitem [{\citenamefont {Kobayashi}\ \emph
  {et~al.}(2025{\natexlab{b}})\citenamefont {Kobayashi}, \citenamefont
  {Otsuka}, \citenamefont {Tanimoto},\ and\ \citenamefont
  {Uchida}}]{Kobayashi:2025ldi}%
  \BibitemOpen
  \bibfield  {author} {\bibinfo {author} {\bibfnamefont {Tatsuo}\ \bibnamefont
  {Kobayashi}}, \bibinfo {author} {\bibfnamefont {Hajime}\ \bibnamefont
  {Otsuka}}, \bibinfo {author} {\bibfnamefont {Morimitsu}\ \bibnamefont
  {Tanimoto}}, \ and\ \bibinfo {author} {\bibfnamefont {Haruki}\ \bibnamefont
  {Uchida}},\ }\bibfield  {title} {\enquote {\bibinfo {title} {{Lepton mass
  textures from non-invertible multiplication rules}},}\ }\href@noop {} {\
  (\bibinfo {year} {2025}{\natexlab{b}})},\ \Eprint
  {http://arxiv.org/abs/2505.07262} {arXiv:2505.07262 [hep-ph]} \BibitemShut
  {NoStop}%
\bibitem [{\citenamefont {Kobayashi}\ \emph
  {et~al.}(2025{\natexlab{c}})\citenamefont {Kobayashi}, \citenamefont
  {Otsuka},\ and\ \citenamefont {Yanagida}}]{Kobayashi:2025thd}%
  \BibitemOpen
  \bibfield  {author} {\bibinfo {author} {\bibfnamefont {Tatsuo}\ \bibnamefont
  {Kobayashi}}, \bibinfo {author} {\bibfnamefont {Hajime}\ \bibnamefont
  {Otsuka}}, \ and\ \bibinfo {author} {\bibfnamefont {Tsutomu~T.}\ \bibnamefont
  {Yanagida}},\ }\bibfield  {title} {\enquote {\bibinfo {title}
  {{Non-invertible Symmetry as a Solution to the Strong CP Problem in a
  GUT-inspired Standard Model}},}\ }\href@noop {} {\  (\bibinfo {year}
  {2025}{\natexlab{c}})},\ \Eprint {http://arxiv.org/abs/2508.12287}
  {arXiv:2508.12287 [hep-ph]} \BibitemShut {NoStop}%
\bibitem [{\citenamefont {Suzuki}\ and\ \citenamefont
  {Xu}(2025)}]{Suzuki:2025oov}%
  \BibitemOpen
  \bibfield  {author} {\bibinfo {author} {\bibfnamefont {Motoo}\ \bibnamefont
  {Suzuki}}\ and\ \bibinfo {author} {\bibfnamefont {Ling-Xiao}\ \bibnamefont
  {Xu}},\ }\bibfield  {title} {\enquote {\bibinfo {title} {{Phenomenological
  implications of a class of non-invertible selection rules}},}\ }\href@noop {}
  {\  (\bibinfo {year} {2025})},\ \Eprint {http://arxiv.org/abs/2503.19964}
  {arXiv:2503.19964 [hep-ph]} \BibitemShut {NoStop}%
\bibitem [{\citenamefont {Kobayashi}\ \emph
  {et~al.}(2025{\natexlab{d}})\citenamefont {Kobayashi}, \citenamefont
  {Okada},\ and\ \citenamefont {Otsuka}}]{Kobayashi:2025cwx}%
  \BibitemOpen
  \bibfield  {author} {\bibinfo {author} {\bibfnamefont {Tatsuo}\ \bibnamefont
  {Kobayashi}}, \bibinfo {author} {\bibfnamefont {Hiroshi}\ \bibnamefont
  {Okada}}, \ and\ \bibinfo {author} {\bibfnamefont {Hajime}\ \bibnamefont
  {Otsuka}},\ }\bibfield  {title} {\enquote {\bibinfo {title} {{Radiative
  neutrino mass models from non-invertible selection rules}},}\ }\href@noop {}
  {\  (\bibinfo {year} {2025}{\natexlab{d}})},\ \Eprint
  {http://arxiv.org/abs/2505.14878} {arXiv:2505.14878 [hep-ph]} \BibitemShut
  {NoStop}%
\bibitem [{\citenamefont {Nomura}\ and\ \citenamefont
  {Okada}(2025)}]{Nomura:2025sod}%
  \BibitemOpen
  \bibfield  {author} {\bibinfo {author} {\bibfnamefont {Takaaki}\ \bibnamefont
  {Nomura}}\ and\ \bibinfo {author} {\bibfnamefont {Hiroshi}\ \bibnamefont
  {Okada}},\ }\bibfield  {title} {\enquote {\bibinfo {title} {{Radiative lepton
  seesaw model in a non-invertible fusion rule and gauged $B-L$ symmetry}},}\
  }\href@noop {} {\  (\bibinfo {year} {2025})},\ \Eprint
  {http://arxiv.org/abs/2506.16706} {arXiv:2506.16706 [hep-ph]} \BibitemShut
  {NoStop}%
\bibitem [{\citenamefont {Nomura}\ and\ \citenamefont
  {Popov}(2025)}]{Nomura:2025yoa}%
  \BibitemOpen
  \bibfield  {author} {\bibinfo {author} {\bibfnamefont {Takaaki}\ \bibnamefont
  {Nomura}}\ and\ \bibinfo {author} {\bibfnamefont {Oleg}\ \bibnamefont
  {Popov}},\ }\bibfield  {title} {\enquote {\bibinfo {title} {{No-group
  Scotogenic Model}},}\ }\href@noop {} {\  (\bibinfo {year} {2025})},\ \Eprint
  {http://arxiv.org/abs/2507.10299} {arXiv:2507.10299 [hep-ph]} \BibitemShut
  {NoStop}%
\bibitem [{\citenamefont {Chen}\ \emph {et~al.}(2025)\citenamefont {Chen},
  \citenamefont {Geng}, \citenamefont {Okada},\ and\ \citenamefont
  {Wu}}]{Chen:2025awz}%
  \BibitemOpen
  \bibfield  {author} {\bibinfo {author} {\bibfnamefont {Jingqian}\
  \bibnamefont {Chen}}, \bibinfo {author} {\bibfnamefont {Chao-Qiang}\
  \bibnamefont {Geng}}, \bibinfo {author} {\bibfnamefont {Hiroshi}\
  \bibnamefont {Okada}}, \ and\ \bibinfo {author} {\bibfnamefont {Jia-Jun}\
  \bibnamefont {Wu}},\ }\bibfield  {title} {\enquote {\bibinfo {title} {{A
  radiative lepton model in a non-invertible fusion rule}},}\ }\href@noop {} {\
   (\bibinfo {year} {2025})},\ \Eprint {http://arxiv.org/abs/2507.11951}
  {arXiv:2507.11951 [hep-ph]} \BibitemShut {NoStop}%
\bibitem [{\citenamefont {Okada}\ and\ \citenamefont
  {Shigekami}(2025)}]{Okada:2025kfm}%
  \BibitemOpen
  \bibfield  {author} {\bibinfo {author} {\bibfnamefont {Hiroshi}\ \bibnamefont
  {Okada}}\ and\ \bibinfo {author} {\bibfnamefont {Yoshihiro}\ \bibnamefont
  {Shigekami}},\ }\bibfield  {title} {\enquote {\bibinfo {title} {{Three-loop
  induced neutrino mass model in a non-invertible symmetry}},}\ }\href@noop {}
  {\  (\bibinfo {year} {2025})},\ \Eprint {http://arxiv.org/abs/2507.16198}
  {arXiv:2507.16198 [hep-ph]} \BibitemShut {NoStop}%
\bibitem [{\citenamefont {Kobayashi}\ \emph
  {et~al.}(2025{\natexlab{e}})\citenamefont {Kobayashi}, \citenamefont {Mita},
  \citenamefont {Otsuka},\ and\ \citenamefont {Sakuma}}]{Kobayashi:2025lar}%
  \BibitemOpen
  \bibfield  {author} {\bibinfo {author} {\bibfnamefont {Tatsuo}\ \bibnamefont
  {Kobayashi}}, \bibinfo {author} {\bibfnamefont {Hironobu}\ \bibnamefont
  {Mita}}, \bibinfo {author} {\bibfnamefont {Hajime}\ \bibnamefont {Otsuka}}, \
  and\ \bibinfo {author} {\bibfnamefont {Riku}\ \bibnamefont {Sakuma}},\
  }\bibfield  {title} {\enquote {\bibinfo {title} {{Matter symmetries in
  supersymmetric standard models from non-invertible selection rules}},}\
  }\href@noop {} {\  (\bibinfo {year} {2025}{\natexlab{e}})},\ \Eprint
  {http://arxiv.org/abs/2506.10241} {arXiv:2506.10241 [hep-ph]} \BibitemShut
  {NoStop}%
\bibitem [{\citenamefont {Dong}\ \emph {et~al.}(2025)\citenamefont {Dong},
  \citenamefont {Jeric}, \citenamefont {Kobayashi}, \citenamefont {Nishida},\
  and\ \citenamefont {Otsuka}}]{Dong:2025jra}%
  \BibitemOpen
  \bibfield  {author} {\bibinfo {author} {\bibfnamefont {Jun}\ \bibnamefont
  {Dong}}, \bibinfo {author} {\bibfnamefont {Tim}\ \bibnamefont {Jeric}},
  \bibinfo {author} {\bibfnamefont {Tatsuo}\ \bibnamefont {Kobayashi}},
  \bibinfo {author} {\bibfnamefont {Ryusei}\ \bibnamefont {Nishida}}, \ and\
  \bibinfo {author} {\bibfnamefont {Hajime}\ \bibnamefont {Otsuka}},\
  }\bibfield  {title} {\enquote {\bibinfo {title} {{On discrete gauging and
  non-invertible selection rules}},}\ }\href@noop {} {\  (\bibinfo {year}
  {2025})},\ \Eprint {http://arxiv.org/abs/2507.02375} {arXiv:2507.02375
  [hep-th]} \BibitemShut {NoStop}%
\bibitem [{\citenamefont {Evans}\ and\ \citenamefont
  {Gannon}(2012)}]{Evans:2012ta}%
  \BibitemOpen
  \bibfield  {author} {\bibinfo {author} {\bibfnamefont {David~E.}\
  \bibnamefont {Evans}}\ and\ \bibinfo {author} {\bibfnamefont {Terry}\
  \bibnamefont {Gannon}},\ }\bibfield  {title} {\enquote {\bibinfo {title}
  {{Near-group fusion categories and their doubles}},}\ }\href@noop {} {\
  (\bibinfo {year} {2012})},\ \Eprint {http://arxiv.org/abs/1208.1500}
  {arXiv:1208.1500 [math.QA]} \BibitemShut {NoStop}%
\bibitem [{\citenamefont {'t~Hooft}(1980)}]{tHooft:1979rat}%
  \BibitemOpen
  \bibfield  {author} {\bibinfo {author} {\bibfnamefont {Gerard}\ \bibnamefont
  {'t~Hooft}},\ }\bibfield  {title} {\enquote {\bibinfo {title} {{Naturalness,
  chiral symmetry, and spontaneous chiral symmetry breaking}},}\ }\href
  {\doibase 10.1007/978-1-4684-7571-5_9} {\bibfield  {journal} {\bibinfo
  {journal} {NATO Sci. Ser. B}\ }\textbf {\bibinfo {volume} {59}},\ \bibinfo
  {pages} {135--157} (\bibinfo {year} {1980})}\BibitemShut {NoStop}%
\bibitem [{\citenamefont {Suzuki}\ and\ \citenamefont
  {Xu}(2026)}]{Suzuki:2025kxz}%
  \BibitemOpen
  \bibfield  {author} {\bibinfo {author} {\bibfnamefont {Motoo}\ \bibnamefont
  {Suzuki}}\ and\ \bibinfo {author} {\bibfnamefont {Ling-Xiao}\ \bibnamefont
  {Xu}},\ }\bibfield  {title} {\enquote {\bibinfo {title} {{Spurion analysis of
  {\ensuremath{\mathbb{Z}}}$_{M}$/{\ensuremath{\mathbb{Z}}}$_{2}$
  non-invertible selection rules: low-order versus all-order zeros}},}\ }\href
  {\doibase 10.1007/JHEP02(2026)227} {\bibfield  {journal} {\bibinfo  {journal}
  {JHEP}\ }\textbf {\bibinfo {volume} {02}},\ \bibinfo {pages} {227} (\bibinfo
  {year} {2026})},\ \Eprint {http://arxiv.org/abs/2510.18972} {arXiv:2510.18972
  [hep-ph]} \BibitemShut {NoStop}%
\bibitem [{\citenamefont {Xu}(2026)}]{Xu:2026nwh}%
  \BibitemOpen
  \bibfield  {author} {\bibinfo {author} {\bibfnamefont {Ling-Xiao}\
  \bibnamefont {Xu}},\ }\bibfield  {title} {\enquote {\bibinfo {title} {{A
  General Prescription for Spurion Analysis of Non-Invertible Selection
  Rules}},}\ }\href@noop {} {\  (\bibinfo {year} {2026})},\ \Eprint
  {http://arxiv.org/abs/2604.09345} {arXiv:2604.09345 [hep-ph]} \BibitemShut
  {NoStop}%
\bibitem [{\citenamefont {Choi}\ \emph {et~al.}(2022)\citenamefont {Choi},
  \citenamefont {Lam},\ and\ \citenamefont {Shao}}]{Choi:2022jqy}%
  \BibitemOpen
  \bibfield  {author} {\bibinfo {author} {\bibfnamefont {Yichul}\ \bibnamefont
  {Choi}}, \bibinfo {author} {\bibfnamefont {Ho~Tat}\ \bibnamefont {Lam}}, \
  and\ \bibinfo {author} {\bibfnamefont {Shu-Heng}\ \bibnamefont {Shao}},\
  }\bibfield  {title} {\enquote {\bibinfo {title} {{Noninvertible Global
  Symmetries in the Standard Model}},}\ }\href {\doibase
  10.1103/PhysRevLett.129.161601} {\bibfield  {journal} {\bibinfo  {journal}
  {Phys. Rev. Lett.}\ }\textbf {\bibinfo {volume} {129}},\ \bibinfo {pages}
  {161601} (\bibinfo {year} {2022})},\ \Eprint
  {http://arxiv.org/abs/2205.05086} {arXiv:2205.05086 [hep-th]} \BibitemShut
  {NoStop}%
\bibitem [{\citenamefont {Cordova}\ and\ \citenamefont
  {Ohmori}(2023)}]{Cordova:2022ieu}%
  \BibitemOpen
  \bibfield  {author} {\bibinfo {author} {\bibfnamefont {Clay}\ \bibnamefont
  {Cordova}}\ and\ \bibinfo {author} {\bibfnamefont {Kantaro}\ \bibnamefont
  {Ohmori}},\ }\bibfield  {title} {\enquote {\bibinfo {title} {{Noninvertible
  Chiral Symmetry and Exponential Hierarchies}},}\ }\href {\doibase
  10.1103/PhysRevX.13.011034} {\bibfield  {journal} {\bibinfo  {journal} {Phys.
  Rev. X}\ }\textbf {\bibinfo {volume} {13}},\ \bibinfo {pages} {011034}
  (\bibinfo {year} {2023})},\ \Eprint {http://arxiv.org/abs/2205.06243}
  {arXiv:2205.06243 [hep-th]} \BibitemShut {NoStop}%
\bibitem [{\citenamefont {Peskin}\ and\ \citenamefont
  {Schroeder}(1995)}]{Peskin:1995ev}%
  \BibitemOpen
  \bibfield  {author} {\bibinfo {author} {\bibfnamefont {Michael~E.}\
  \bibnamefont {Peskin}}\ and\ \bibinfo {author} {\bibfnamefont {Daniel~V.}\
  \bibnamefont {Schroeder}},\ }\href {\doibase 10.1201/9780429503559} {\emph
  {\bibinfo {title} {{An Introduction to quantum field theory}}}}\ (\bibinfo
  {publisher} {Addison-Wesley},\ \bibinfo {address} {Reading, USA},\ \bibinfo
  {year} {1995})\BibitemShut {NoStop}%
\bibitem [{\citenamefont {Weinberg}(1968)}]{Weinberg:1968de}%
  \BibitemOpen
  \bibfield  {author} {\bibinfo {author} {\bibfnamefont {Steven}\ \bibnamefont
  {Weinberg}},\ }\bibfield  {title} {\enquote {\bibinfo {title} {{Nonlinear
  realizations of chiral symmetry}},}\ }\href {\doibase
  10.1103/PhysRev.166.1568} {\bibfield  {journal} {\bibinfo  {journal} {Phys.
  Rev.}\ }\textbf {\bibinfo {volume} {166}},\ \bibinfo {pages} {1568--1577}
  (\bibinfo {year} {1968})}\BibitemShut {NoStop}%
\bibitem [{\citenamefont {D'Ambrosio}\ \emph {et~al.}(2002)\citenamefont
  {D'Ambrosio}, \citenamefont {Giudice}, \citenamefont {Isidori},\ and\
  \citenamefont {Strumia}}]{DAmbrosio:2002vsn}%
  \BibitemOpen
  \bibfield  {author} {\bibinfo {author} {\bibfnamefont {G.}~\bibnamefont
  {D'Ambrosio}}, \bibinfo {author} {\bibfnamefont {G.~F.}\ \bibnamefont
  {Giudice}}, \bibinfo {author} {\bibfnamefont {G.}~\bibnamefont {Isidori}}, \
  and\ \bibinfo {author} {\bibfnamefont {A.}~\bibnamefont {Strumia}},\
  }\bibfield  {title} {\enquote {\bibinfo {title} {{Minimal flavor violation:
  An Effective field theory approach}},}\ }\href {\doibase
  10.1016/S0550-3213(02)00836-2} {\bibfield  {journal} {\bibinfo  {journal}
  {Nucl. Phys. B}\ }\textbf {\bibinfo {volume} {645}},\ \bibinfo {pages}
  {155--187} (\bibinfo {year} {2002})},\ \Eprint
  {http://arxiv.org/abs/hep-ph/0207036} {arXiv:hep-ph/0207036} \BibitemShut
  {NoStop}%
\bibitem [{\citenamefont {Grinstein}\ \emph {et~al.}(2024)\citenamefont
  {Grinstein}, \citenamefont {Lu}, \citenamefont {Mir\'o},\ and\ \citenamefont
  {Qu\'\i{}lez}}]{Grinstein:2024iyf}%
  \BibitemOpen
  \bibfield  {author} {\bibinfo {author} {\bibfnamefont {Benjam\'\i{}n}\
  \bibnamefont {Grinstein}}, \bibinfo {author} {\bibfnamefont {Xiaochuan}\
  \bibnamefont {Lu}}, \bibinfo {author} {\bibfnamefont {Carlos}\ \bibnamefont
  {Mir\'o}}, \ and\ \bibinfo {author} {\bibfnamefont {Pablo}\ \bibnamefont
  {Qu\'\i{}lez}},\ }\bibfield  {title} {\enquote {\bibinfo {title} {{Most
  general EFTs from spurion analysis: Hilbert series and Minimal Lepton Flavor
  Violation}},}\ }\href@noop {} {\  (\bibinfo {year} {2024})},\ \Eprint
  {http://arxiv.org/abs/2412.16285} {arXiv:2412.16285 [hep-ph]} \BibitemShut
  {NoStop}%
\end{thebibliography}%

\end{document}